\pdfoutput=1

\documentclass[11pt]{article}
\usepackage{jheppub}

\usepackage{amsmath}
\usepackage{amssymb}
\usepackage{graphicx,color,slashed}
\usepackage{ifpdf}
\usepackage[english]{babel}
\usepackage{url}

\usepackage{float}

\allowdisplaybreaks[1]

\font\tenshuffle=shuffle10 \font\sevenshuffle=shuffle7 \font\fiveshuffle=shuffle7 at 5pt
\def\shuffle{{%
\def\Dshuffle{\mathbin{\hbox{\tenshuffle\char'001}}}%
\def\Sshuffle{\mathbin{\hbox{\sevenshuffle\char'001}}}%
\def\SSshuffle{\mathbin{\hbox{\fiveshuffle\char'001}}}%
\mathchoice{\Dshuffle}{\Dshuffle}{\Sshuffle}{\SSshuffle}}}

\restylefloat{figure}
\definecolor{dgreen}{rgb}{0,0.70,0.30}
\definecolor{gold}{rgb}{0.85,.66,0}
\definecolor{purple}{rgb}{1.0,0.3,0.6}

\def\dz{{\rm d}z}

\def\be{\begin{equation}}
\def\ee{\end{equation}}
\def\ba{\begin{array}}
\def\ea{\end{array}}

\newcommand{\bea}{\begin{eqnarray}}
\newcommand{\eea}{\end{eqnarray}}
\newcommand{\Tr}{{\rm Tr}}
\newcommand\myatop[2]{\genfrac{}{}{0pt}{}{#1}{#2}}

\def\beq{\begin{equation}}
\def\eeq{\end{equation}}
\def\Re{{\rm Re\,}}

\newcommand{\co}{\ , \ \ \ \ \ \ }
\newcommand{\dd}{\mathrm{d}}
\newcommand{\te}{\textrm}
\newcommand{\ap}{{\alpha'}}

\title{Semi-abelian Z-theory: NLSM+$\phi^3$ from the open string}
\author[a]{John Joseph M. Carrasco,}
\author[b]{Carlos R. Mafra,}
\author[c]{Oliver Schlotterer}

\affiliation[a]{Institut de Physique Th\'eorique,
CEA--Saclay, F--91191 Gif-sur-Yvette cedex, France}
\affiliation[b]{STAG Research Centre and Mathematical Sciences,
University of Southampton, UK}
\affiliation[c]{Max--Planck--Institut f\"ur Gravitationsphysik,
Albert--Einstein--Institut,
14476 Potsdam, Germany}
\emailAdd{john-joseph.carrasco@cea.fr}
\emailAdd{c.r.mafra@soton.ac.uk}
\emailAdd{olivers@aei.mpg.de}

\date{\today}

\abstract{We continue our investigation of $Z$-theory, the second double-copy component 
of open-string tree-level interactions besides super-Yang--Mills (sYM). We show that the 
amplitudes of the extended non-linear sigma model (NLSM) recently considered by Cachazo, 
Cha, and Mizera are reproduced by the leading $\ap$-order of $Z$-theory amplitudes in the
semi-abelian case. The extension refers to a coupling of NLSM pions to bi-adjoint scalars, and
the semi-abelian case involves to a {\it partial} symmetrization over one of the color orderings 
that characterize the $Z$-theory amplitudes. Alternatively, the partial symmetrization corresponds 
to a mixed interaction among abelian and non-abelian states in the underlying open-superstring 
amplitude. We simplify these permutation sums via monodromy relations which greatly increase 
the efficiency in extracting the $\ap$-expansion of these amplitudes. Their $\ap$-corrections encode 
higher-derivative interactions between NLSM pions and bi-colored scalars all of which obey the 
duality between color and kinematics.  Through double-copy, these results can be used to generate 
the predictions of supersymmetric Dirac--Born--Infeld--Volkov--Akulov theory coupled with sYM 
as well as a complete tower of higher-order $\ap$-corrections.
}

\begin{document}

\maketitle{}

\setcounter{tocdepth}{2}

\numberwithin{equation}{section}

\newpage

\section{Introduction}

$Z$-theory~\cite{abZtheory, nonabZ} refers to the $\ap$ dependent  theory of bi-colored\footnote{It may be tempting to refer to the $Z$-theory scalars as ``bi-adjoint'', since in the low-energy ($\ap\rightarrow0$) limit, non-abelian $Z$-theory becomes bi-adjoint $\phi^3$.  We make a different choice here to emphasize the following important point: the $\ap$-corrections imply that $Z$-theory scalars are not trivially Lie-algebra valued w.r.t.\ one of the gauge groups -- the one dressed by string Chan--Paton factors.  Charges under the gauge groups are referred to as ``color'' throughout this work which may equivalently be replaced by ``flavour''.}  
scalars whose double copy~\cite{Kawai:1985xq,BCJ,Cachazo:2013gna} with maximally supersymmetric
Yang--Mills theory (sYM)~\cite{Brink:1976bc} generates the tree-level scattering predictions of
the open superstring. $Z$-theory was originally defined 
 by taking its amplitudes to be the set of doubly-ordered functions $Z_\sigma(\tau)$ of ref.~\cite{Zfunctions}
-- iterated integrals over the boundary of a worldsheet of disk
topology -- which arise in the tree-level amplitudes of the open
superstring \cite{Mafra:2011nv,Mafra:2011nw}. The
complete $\ap$-expansion of the non-linear $Z$-theory equations of
motion is  pinpointed in ref.~\cite{nonabZ}.

To translate these doubly-ordered $Z_\sigma(\tau)$-functions to field-theory
scattering amplitudes, one dresses the permutation $\sigma \in S_n$ encoding the integration domain with
the Chan--Paton (CP) factors associated with open-string endpoints.
Depending on whether the CP
factors are entirely non-abelian or abelian, the low-energy limits of the corresponding
$Z$-theory amplitudes reproduce the tree-level interactions of either bi-adjoint scalar
particles~\cite{Bern:1999bx, Cachazo:2013iea}, or non-linear sigma model (NLSM)
pions\footnote{See \cite{Cronin:1967jq, Weinberg:1966fm,Weinberg:1968de,Brown:1967qh,Chang:1967zza} and \cite{
Susskind:1970gf,Osborn:1969ku,Ellis:1970nt,Kampf:2013vha} for earlier references on the NLSM and its tree-level amplitudes, respectively.}~\cite{abZtheory}.   $Z$-theory amplitudes offer a fascinating laboratory to study 
stringy emergence in a new and technically much simpler context. From a double-copy perspective they  isolate, in 
a {\em scalar} field theory, what is ultra UV-soft\footnote{For discussions of the famously exponentially suppressed UV behaviour of string  scattering see e.g. refs.~\cite{Gross:1987kza,Caron-Huot:2016icg,Arkani-HamedStrings2016}.} in higher-derivative tree-level predictions of the open superstring. 

The main result of this work concerns the semi-abelian
version of $Z$-amplitudes -- those
involving a mixture of abelian and non-abelian CP factors. Their low-energy theory
will be identified with interactions among NLSM pions and bi-adjoint scalar particles (NLSM$+\phi^3$).
Amplitudes in this theory have been recently studied \cite{Cachazo:2016njl} in a 
Cachazo--He--Yuan (CHY) representation \cite{Cachazo:2013gna, Cachazo:2013hca, Cachazo:2013iea}.  
To be concrete, we will generalize the emergence of color-stripped NLSM amplitudes
from completely abelianized disk integrals or abelian $Z$-theory amplitudes \cite{abZtheory},
 \beq
 A^{\rm NLSM}(1,2,\ldots,n) = \lim_{\ap \rightarrow 0} \frac{1}{n \ap}
 \int _{\mathbb R^n} \frac{ \dd z_1 \, \dd z_2 \, \ldots \, \dd z_n   }
 { {\rm vol}(SL(2,{ \mathbb R} )) }
 \frac{\prod^n_{i<j} |z_{ij}|^{\ap k_{i}\cdot k_{j}} }{ z_{12}z_{23} \ldots z_{n-1,n} z_{n1}} \ ,
 \label{NLSM}
 \eeq
where $z_{ij} \equiv z_i{-}z_j$. In close analogy to (\ref{NLSM}), we will identify the doubly-stripped amplitudes of the 
(NLSM+$\phi^3$) theory in the low-energy limit of semi-abelian
$Z$-amplitudes (\ref{semiab02}), 
 \beq
 A^{{\rm NLSM} + \phi^3}(1,2,\ldots,r|\tau(1,2,\ldots,n)) = \lim_{\ap \rightarrow 0} 
 \ap^{r-3-\delta} \! \! \!
  \! \! \! \! \! \! \! \! \! \! \! \! \! \! \! \! \!  \int \limits_{-\infty \leq  z_{1} \leq z_{2}\leq \ldots \leq z_{r} \leq \infty }  \! \! \! \! \! \! \! \! \! \! \! \! \! \! \! \!
  \frac{ \dd z_1 \, \dd z_2 \, \ldots \, \dd z_n   }{ {\rm vol}(SL(2, {\mathbb R})) } \frac{\prod^n_{i<j} |z_{ij}|^{\ap k_{i}\cdot k_{j}} }{\tau( z_{12}z_{23} \ldots z_{n-1,n} z_{n1} )} \ ,
 \label{NLSMphi3}
 \eeq
with $2{\leq} r{\leq} n$ external bi-adjoint scalars. The power of $\ap$ is $\delta =0$ and $\delta=1$ for even and odd numbers of pions $n{-}r$, respectively, and the second ordering referring to the integrand is governed by a permutation $\tau \in S_n$.

As first realized in ref.~\cite{Cachazo:2014xea}, the NLSM double copies with sYM to generate predictions in Dirac--Born--Infeld--Volkov--Akulov theory (DBIVA)~--- the supersymmetric completion of Born--Infeld (see e.g.~\cite{Bergshoeff:2013pia}). The abelianized open string, as an all-order double copy of abelian $Z$-theory with sYM, provides $\ap$-corrections to DBIVA~\cite{abZtheory}.  Similarly, as the NLSM$+ \phi^3$  double copies  with sYM to generate predictions in  DBIVA coupled with sYM, the semi-abelian open string provides an all-order $\ap$-completion to  DBIVA+sYM.  One can either use the field-theory ($\ap\to0$) Kawai--Lewellen--Tye (KLT) type-relations at tree level \cite{Kawai:1985xq}, often encoded in a momentum kernel~\cite{momentumKernel}, to double copy ordered amplitudes, or solve for Jacobi-satisfying numerators and take the double copy graph by graph, following the duality between color and kinematics due to Bern, Johansson and one of the current authors (BCJ) \cite{BCJ,loopBCJ}.  As sYM amplitudes are by now standard textbook material~\cite{Elvang:2013cua}, the new ingredient which we provide  is the understanding of how to generate the various $\ap$-components of semi-abelian $Z$-theory.  Additionally, with these amplitudes in hand, one can even consider the double copy of semi-abelian $Z$-theory with itself, which results in a set of higher-derivative corrections to the theory of special Galileons coupled with NLSM$+ \phi^3$ discussed in ref.\ \cite{Cachazo:2016njl}. 

It should be noted that the type of non-linear symmetry at work in effective field theories like the NLSM and (through double copy with sYM) in DBIVA, has recently garnered some attention from applications to cosmology. Volkov--Akulov-type constrained ${{\cal N}=1}$ superfields allow for technically simple inflationary models~\cite{Ferrara:2014kva, Ferrara:2015tyn, Carrasco:2015pla,Carrasco:2015iij} and descriptions of dark energy \cite{Bergshoeff:2015tra,Hasegawa:2015bza, Kuzenko:2015yxa, Bandos:2015xnf}. This, as well as independent advances in the notion of a soft bootstrap, has motivated renewed interest in understanding the effect of such non-linear symmetries on the S-matrix, with special attention to its soft limits\footnote{The ability of soft limits of a theory's S-matrix to encode its symmetries has long been appreciated, from the conception of what became known as Adler zeros \cite{Adler:1964um}, to the imprint of coset symmetry on double-soft limits~\cite{ArkaniHamed:2008gz}.}, see e.g.~\cite{Kallosh:2016qvo, Kallosh:2016lwj,Cheung:2016drk, Du:2016njc} and references therein. It should be interesting to discover what symmetries survive, and indeed emerge from, the higher-order string-theory type completion encoded in the $Z$-theory amplitudes presented in \cite{abZtheory} and here.

To understand the functions at work we will recall in Section
\ref{sec:Review} the definition of the disk integrals or
$Z$-amplitudes at the heart of the CP-stripped open string. 
As explained in section \ref{sect3}, the semi-abelian case requires
partial symmetrizations over CP orderings which are simplified using
monodromy relations \cite{BjerrumBohr:2009rd, Stieberger:2009hq,
goodCDOpenString}. These techniques for evaluating (\ref{NLSMphi3})
together with the Berends--Giele recursion for non-abelian $Z$-theory
amplitudes \cite{nonabZ} give efficient access to the
higher-derivative interactions between pions and bi-colored scalars.
Integration-by-parts relations among the disk integrals guarantee
that the BCJ duality between color and kinematics \cite{BCJ,loopBCJ}  
holds to all orders in $\ap$.

In the low-energy limit of semi-abelian $Z$-amplitudes (\ref{NLSMphi3}) detailed in section
\ref{sect4} we will make contact with recent results involving
NLSM$+\phi^3$ \cite{Cachazo:2016njl}. $Z$-theory finds exact agreement 
with the tree amplitudes of ref.~\cite{Cachazo:2016njl} for an even number of pions, while
yielding additional couplings for odd numbers
of pions. Our low-energy results reveal novel amplitude relations
between the extended NLSM and pure $\phi^3$ theory and imply
simplifications of their CHY description \cite{Cachazo:2014xea,
Cachazo:2016njl}.

\section{Review}
\label{sec:Review}

He we provide a lightening overview of doubly-ordered $Z$-theory
amplitudes so as to set up the main results. We refer the reader
to~\cite{abZtheory, nonabZ} for detailed
reviews of $Z$-theory as well as properties of color-kinematics and the 
double copy.

As discussed in \cite{abZtheory}, it is possible, and indeed quite
intriguing, to  interpret the iterated disk integrals of the
CP stripped open-string amplitude as predictions in an effective
field theory.  First we define the {\it doubly-ordered}
$Z$-functions~\cite{Zfunctions},
\be
\label{sig04}
Z_{\sigma(1,2,\ldots,n)}(\tau(1,2,\ldots,n)) \equiv \ap^{n-3}\! \! \! \! \! \! \! \! \! \! \! \! \! \! \! \! \! \! \! \! \!  
\int \limits_{ -\infty \leq  z_{\sigma(1)} \leq z_{\sigma(2)}\leq \ldots \leq z_{\sigma(n)} \leq \infty } \! \! \! \! \! \! \! \! \! \! \! \! \! \! \! \! \! \! \! \! \!  {\dz_1 \ \dz_2 \ \cdots  \ \dz_n \over {\rm vol}(SL(2,{\mathbb  R}))}
 { \prod_{i<j}^n |z_{ij}|^{\ap s_{ij}}  \over \tau(z_{1 2} z_{2 3} \ldots z_{{n-1},n} z_{n1})}\,,
\ee
with permutations $\sigma,\tau \in S_n$.
The field-theory ordering $\tau$ determines the cyclic product
of inverse $z_{ij}\equiv z_i - z_j$ in the integrand, and
integration-by-parts manipulations imply~\cite{Zfunctions} that
different choices of $\tau$ are related by the {\em field-theory} BCJ relations \cite{BCJ}
\be
\sum_{j=2}^{n-1}  (k_1\cdot k_{23\ldots j} )\,
Z_{\sigma(1,2,\ldots,n)}(2,3,\ldots, j ,1,j{+}1,\ldots ,n)= 0
\label{BCJ}
\ee
at fixed $\sigma$. The CP-ordering $\sigma$, on the other hand,
constrains the domain of integration such that $z_{\sigma(i)}\leq z_{\sigma(i+1)}$
for $i=1,2,\ldots,n{-}1$, where $Z_{12\ldots n}(\ldots)$ is cyclically equivalent to 
$Z_{2\ldots n1}(\ldots)$. The $\ap$-dependent {\em string-theory} monodromy
relations~\cite{BjerrumBohr:2009rd, Stieberger:2009hq}
\be
\sum_{j=1}^{n-1} e^{i\pi \ap k_1\cdot k_{23\ldots j} }\,  Z_{23\ldots j ,1,j+1\ldots n}(\tau(1,2,\ldots,n))  = 0
\label{monodromy}
\ee
intertwine the contributions from different integration domains
resulting ultimately in an $(n{-}3)!$-basis at fixed integrand
ordering $\tau$. Accordingly, the $\sigma$-ordering in (\ref{sig04})
will also be referred to as the monodromy ordering. As we reserve the right 
to dress both orderings with color-information, we will distinguish the monodromy
related dressings as CP factors, and the field-theory order relevant dressings
as color-factors.  Note that our conventions for Mandelstam invariants in (\ref{sig04}) and
multiparticle momenta in (\ref{monodromy}) are fixed by
\be
k_{12\ldots p} \equiv k_1+k_2+\ldots+k_p \ , \ \ \ \ \ \ 
s_{12\ldots p} \equiv \frac{1}{2} \, k_{12\ldots p}^2 = \sum_{i<j}^p k_i \cdot k_j \ .
\label{conv}
\ee
The prefactor $\ap^{n-3}$ in (\ref{sig04}) is designed to obtain the doubly-partial
amplitudes of the bi-adjoint scalar theory $m[\cdot | \cdot ]$ in the limit \cite{Cachazo:2013iea}
\beq
\lim_{\ap \rightarrow 0 } Z_{\sigma(1,2,\ldots,n)}(\tau(1,2,\ldots,n)) = m[\sigma(1,2,\ldots,n) \, | \, \tau(1,2,\ldots,n)]  \, ,
 \label{ftlimX}
\eeq
see  \cite{Mafra:2016ltu} and \cite{nonabZ} for Berends--Giele recursions for the field-theory amplitudes $m[\cdot | \cdot ]$ and the full-fledged disk integrals (\ref{sig04}), respectively.

Perhaps the most natural way to think about $Z$-theory as an effective field
theory is as a doubly-colored scalar theory where one color
(corresponding to color order $\sigma$, whose generators we will
annotate with $t^a$) is provided by the stringy\footnote{Of course
there is nothing stringy about the CP factors themselves, rather the
doubly-ordered amplitude obeys the string monodromy relations on the
order dressed by the CP factors.} CP factors. The CP color mixes
with all higher-order kinetic
terms\footnote{Indeed these higher-derivative terms are
responsible for the CP ordering satisfying monodromy relations as
opposed to the field-theory relations of the field-theory
color-ordering.} depending on $\ap\, k_i \cdot k_j$. The other color (corresponding to color order
$\tau$, whose generators we will annotate with $T^a$) represents a
familiar field-theory non-abelian color dressing.  

As mentioned in the introduction, to achieve familiar color-ordered
amplitudes we must dress the doubly-ordered $Z_\sigma(\tau)$ along
one of their orderings.  Dressing $\sigma$ with the CP factors leaves us with a
manifestly factorizable theory whose amplitudes obey the standard
field-theory BCJ relations (\ref{BCJ}). Explicitly, we sum (\ref{sig04}) over all
distinct $\sigma$ orders, weighting each $Z_{\sigma}(\tau)$ with the
$\sigma$-ordered CP trace:
\be
{\cal Z}\left(\tau(1,2,\ldots,n)\right) \equiv \sum_{\sigma \in S_{n-1}} \Tr(t^1t^{\sigma(2)}\cdots t^{\sigma(n)}) Z_{1,\sigma(2,\ldots ,n)}(\tau (1,2,\ldots,n))  \,.
\label{eqn:masterZ}
\ee
Starting from the CP-dressed $Z$-theory amplitude
(\ref{eqn:masterZ}), the color-dressed open-string amplitude
\cite{Mafra:2011nv} can be written in the form \cite{Zfunctions}
\begin{align}
&M^{\rm open}_n =  \sum_{\tau,\rho \in S_{n-3}} 
{\cal Z}\left(1,\tau(2,\ldots,n{-}2),n,n{-}1\right) \label{openKLT}\\
& \ \ \ \times
S[ \tau(23\ldots n{-}2) \,
| \, \rho(23\ldots n{-}2) 
]_1 \, A^{\rm YM}(1,\rho(2,\ldots,n{-}2),n{-}1,n) 
\notag
\end{align}
of the KLT relations for supergravity
amplitudes \cite{Kawai:1985xq, Bern:1998sv}. The matrix $S[\cdot |\cdot ]_1$
is known as the field-theory momentum kernel \cite{momentumKernel} and allows for the recursive
representation \cite{abZtheory},
\beq
S[A,j\, | \, B,j,C]_i =
(k_{iB}\cdot k_{j}) S[A\, | \, B,C]_i
 \co S[\emptyset \, | \, \emptyset]_i \equiv 1  \ ,
\label{momk}
\eeq
with multiparticle labels such as $B=(b_1,\ldots,b_{p})$ and
$C=(c_1,\ldots,c_{q})$, multiparticle momentum $k_{iB}\equiv k_i +
k_{b_1} + \cdots + k_{b_p}$ and composite label
$B,C=(b_1,\ldots,b_{p},c_1,\ldots,c_q)$. In the next section, we will derive
simplified representations for the CP dressed $Z$-amplitudes (\ref{eqn:masterZ}) 
when some of the generators $t^a$ are abelian. In this semi-abelian case, the open-string amplitudes
(\ref{openKLT}) encode a UV completion of supersymmetric DBIVA coupled with sYM \cite{Metsaev:1987qp}, 
and our subsequent results on ${\cal Z}(\ldots)$ should offer insight into the structure of its tree-level S-matrix.


\section{Semi-abelian $Z$-theory amplitudes}
\label{sect3}

\subsection{A structural perspective}

In the case of some abelian CP-charged particles where
$t^a\to{\mathbf 1}$, the traces in (\ref{eqn:masterZ}) reduce to
only the relevant non-abelian generators. If there are $r$
non-abelian charged particles with labels $1,2,\ldots, r$ and $n{-}r$
abelian particles, the
color-ordered CP-dressed ${\cal Z}(\tau)$ amplitude  \eqref{eqn:masterZ} can be written as
\be
{\cal Z}\left(\tau(1,2,\ldots,n)\right) \, \big|_{t^{r+1},\ldots,t^n \rightarrow {\mathbf 1}} = \sum_{\sigma \in S_{r-1}} \Tr(t^1t^{\sigma(2)}\cdots t^{\sigma(r)}) {Z}_{1,\sigma(2,3,\ldots,r)}(\tau(1,2,\ldots,n))  \,.
\label{eqn:sabZ}
\ee
In the notation
 $\Sigma(1,2,\ldots,r)\equiv\{1,\sigma(2,3,\ldots,r)\}$ for their
 integration domain, the semi-abelianized doubly-ordered
 ${Z}_\Sigma(\tau)$-amplitudes with $r\leq n$ are given as
  \beq
 {Z}_{\Sigma(12\ldots r)}\left(\tau(1,2,\ldots,n)\right) \equiv \ap^{n-3} \! \! \! \! \! \! \! \! \! \! \! \! \! \! \! \! \! \! \! \! \!  \int \limits_{-\infty \leq  z_{\Sigma(1)} \leq z_{\Sigma(2)}\leq \ldots \leq z_{\Sigma(r)} \leq \infty }\! \! \! \! \! \! \! \! \! \! \! \! \! \! \! \! \! \! \! \! \frac{ \dd z_1 \, \dd z_2 \, \ldots \, \dd z_n   }{ {\rm vol}(SL(2,{\mathbb R})) } \frac{\prod^n_{i<j} |z_{ij}|^{\ap s_{ij}} }{\tau(z_{12}z_{23} \ldots z_{n-1,n} z_{n1})} \ ,
\label{semiab02}
\eeq
where the punctures $z_{r+1},\ldots,z_n$ are understood to be integrated over the range $(z_{\Sigma(1)},\infty)$.
Note that we have $\Sigma \equiv \{ \emptyset \}$ for the abelianized $Z$-theory 
introduced in \cite{abZtheory}, and so what would putatively be a
doubly-ordered integral becomes the only single-ordered integral
relevant to the theory at a given multiplicity, with the order $\tau_n \equiv
\tau(1,2,\ldots,n)$.

Both the monodromy relations ~\cite{BjerrumBohr:2009rd,
Stieberger:2009hq} and the recent all-multiplicity developments on
$\alpha'$-expansions \cite{Schlotterer:2012ny, Zfunctions,
Broedel:2013aza, nonabZ} are tailored to non-abelian disk integrals
$Z_{\rho}(\tau)$ in (\ref{sig04}), where $\rho$ and $\tau$ refer to
all the $n$ particles. In order to export these results to the
semi-abelian disk integrals of (\ref{eqn:sabZ}), the latter need to
be expressed in terms of their completely ordered counterparts
$Z_{\rho}(\tau)$. 

Of course, the inequalities among $z_1,z_2,\ldots,z_r$ imposed by
the $\Sigma$-ordering in (\ref{semiab02}) can always be translated
into a combination of $n$-particle orderings,
\beq
{Z}_{1,\sigma(2,3,\ldots,r)}(\tau_n)  = \sum_{\myatop{\rho(2,\ldots,n) \in \big[\sigma(2,\ldots,r) \shuffle}{ \,r{+}1\, \shuffle \,r{+}2 \,\shuffle \ldots \shuffle\, n\big]}}
{Z}_{1,\rho(2,3,\ldots,n)}(\tau_n)  \ ,
\label{semishuffle}
\eeq
where the shuffle symbol acting on words $B=(b_1,\ldots,b_{p})$ and
$C=(c_1,\ldots,c_{q})$ can be recursively defined by
\beq
\emptyset\shuffle B = B\shuffle\emptyset = B \ , \ \ \ \ \ \ 
B\shuffle C \equiv b_1(b_2 \ldots b_{p} \shuffle C) + c_1(c_2 \ldots c_{q}
\shuffle B)\,.
\eeq
However, this ``naive'' expansion of semi-abelian $Z$-amplitudes
$Z_{\Sigma(12\ldots r)}(\tau_n)= {Z}_{1,\sigma(2,\ldots,r)}(\tau_n)$ in terms of their non-abelian counterparts
$Z_{\rho}(\tau)$ usually carries a lot of redundancies and obscures
the leading low-energy order. Hence, we will be interested in a
simplified representation in terms of $(n{-}2)!$ non-abelian
orderings $\rho=\rho(2,3,\ldots,n{-}1)$ which is specified by an $\ap$-dependent
coefficient matrix ${\cal W}_\ap(\Sigma\, |\,\rho)$,
\be
Z_\Sigma(\tau_n) \equiv \sum_{\rho \in S_{n-2}} {\cal W}_\ap(\Sigma \, | \, \rho(2,3,\ldots,n{-}1)) \, Z_{1,\rho(2,3,\ldots,n-1),n}(\tau_n)\,,
\label{structural}
\ee
The expansion coefficients in the matrix ${\cal
W}_\ap(\Sigma\,|\,\rho)$ will be identified as trigonometric
functions of $\ap s_{ij}$ universal to all $\tau_n$ which clarify
the first non-vanishing order of $\ap$. This approach will be seen
to yield particularly useful expressions for $Z_{\Sigma}(\tau) $ with a small number $r$
of non-abelian CP factors, to expose their leading low-energy
order, to simplify the identification of their field-theory limit
and to render the computation of their $\ap$-expansion more
efficient.

The desired form (\ref{structural}) of semi-abelian $Z$-theory
amplitudes can be achieved by exploiting the monodromy relations at
the level of the CP-dressed integrals (\ref{eqn:masterZ})
\cite{goodCDOpenString},
\be
{\cal Z}(\tau_n) =  \sum_{\sigma \in S_{n-2}} \Tr( [[\cdots [[t^1,t^{\sigma(2)}]_\ap, t^{\sigma(3)}]_\ap,\cdots]_\ap,t^{\sigma({n-1})}]_\ap ~ t^{n})\, Z_{1,\sigma(2,3,\ldots,n-1), n}(\tau_n)  \, \ .
\label{CPdressed}
\ee
In the context of the color-dressed open superstring
(\ref{openKLT}), this can be viewed\footnote{Note that with $\ap\to0$ the trace of commutators is exactly the color weight of the appropriate half-ladder diagram.} as a generalization of the
Del-Duca--Dixon--Maltoni representation of color-dressed sYM
amplitudes \cite{DelDuca:1999rs}. The complex phases seen in the
monodromy relations (\ref{monodromy}) are absorbed into the
symmetric version of the $\ap$-weighted commutator
of~\cite{goodCDOpenString},
\begin{align}
[t^{i_1} t^{i_2} \ldots t^{i_p} , t^{j_1} t^{j_2} \ldots t^{j_q} ]_{\alpha'}
&\equiv e^{i x_{i_1 i_2\ldots i_p , j_1 j_2\ldots j_q} }\,
(t^{i_1} t^{i_2} \ldots t^{i_p})\,( t^{j_1} t^{j_2} \ldots t^{j_q}) \label{eqn:commRelns} \\
 &- e^{-i x_{i_1 i_2\ldots i_p , j_1 j_2 \ldots j_q} }
\,(t^{j_1} t^{j_2} \ldots t^{j_q} )\,(t^{i_1} t^{i_2} \ldots t^{i_p}) \, , \notag
\end{align}
where the exponents are furnished by rescaled Mandelstam invariants (\ref{conv})
\beq
x_{i_1 i_2\ldots i_p , j_1 j_2 \ldots j_q}  \equiv \frac{\pi \alpha'}{2} \, k_{i_1 i_2\ldots i_p}\cdot k_{ j_1 j_2 \ldots j_q} \ .
\label{semiab53}
\eeq
A simplified representation (\ref{structural}) of semi-abelian $Z$-theory amplitudes
(\ref{semiab02}), in particular the explicit form of the coefficient matrix
${\cal W}_\ap(\Sigma \, | \, \rho) $ for $\Sigma \equiv \Sigma(12\ldots r)$, follows by isolating the
coefficient of a given CP trace in (\ref{CPdressed}) after abelianizing
$t^{ r+1},t^{ r+2},\ldots,t^{n} \rightarrow \mathbf{1}$.

\subsection{Simplified representation of abelian $Z$-theory amplitudes}

Once we specialize (\ref{CPdressed}) to abelian gauge bosons with
$t^j \rightarrow {\mathbf 1}$ for $j=1,2,\ldots,n$, the
$\ap$-weighted commutators (\ref{eqn:commRelns}) reduce to
sine-functions and yield the following simplified expression for
the abelian $Z$-theory amplitudes of \cite{abZtheory},
\beq
{\cal{Z}}(\tau_n) \, \big|_{t^j \rightarrow {\mathbf 1}} \equiv Z_\times(\tau_n) =   (2 i)^{n-2} \sum_{\rho \in S_{n-2}}  Z_{1,\rho(2,3,\ldots,n-1),n} (\tau_n)  \prod_{k=2}^{n-1} 
 \sin(x_{1\rho(23\cdots(k-1)),\rho(k)}) \, .
\label{eqn:abelianZ}
\eeq
We continue to use the shorthand $\tau_n \equiv \tau(1,2,\ldots,n)$
for the integrands, and by the vanishing of odd-multiplicity
instances $Z_\times(\tau_{2m-1}) =0$, the multiplicity $n$ is taken
to be even, e.g.
\begin{align}
Z_{\times}(\tau_4) &= 4 \sin^2 \Big( \frac{ \pi \alpha'}{2} \, s_{12} \Big) \, Z_{1234}(\tau_4) + 4\sin^2 \Big( \frac{ \pi \alpha'}{2} \, s_{13} \Big) \, Z_{1324}(\tau_4)
\label{semiab56a} \\
Z_{\times}(\tau_6) &=16 \sum_{\rho \in S_4}   \sin \Big( \frac{ \pi \alpha'}{2} \, s_{1\rho(2)} \Big)\,  \sin \Big( \frac{ \pi \alpha'}{2} \, (s_{1\rho(3)}+s_{\rho(23)}) \Big) \label{semiab56b} \\
& \ \ \ \ \ \ \ \ \times 
\,  \sin \Big( \frac{ \pi \alpha'}{2} \, (s_{\rho(45)}+s_{\rho(4)6}) \Big)
\,  \sin \Big( \frac{ \pi \alpha'}{2} \, s_{\rho(5)6} \Big)  Z_{1\rho(2345)6}(\tau_6) \ . \notag
\end{align}
Given that each factor of
\beq
\sin(x_{12\ldots j-1,j}) = \sin \Big( \frac{ \pi \alpha'}{2} \, k_{12\ldots j-1}\cdot k_j \Big) = \frac{ \pi \alpha'}{2} \, k_{12\ldots j-1}\cdot k_j + {\cal O}(\ap^3)
\label{semiab57}
\eeq
introduces one power of $\pi \alpha' k^2$ into the low-energy limit, the leading behaviour of
\beq
Z_{\times}(\tau(1,2,\ldots,n)) = {\cal O} (\ap^{(n-2)} )
\label{semiab58}
\eeq
is manifest in (\ref{eqn:abelianZ}), in lines with the
identification of the NLSM amplitude in \cite{abZtheory}. Hence, the
sine-factors bypass the $\ap$-expansion of disk integrals
$Z_{1,\rho,n}(\tau)$ to the order $\ap^{n-2}$ when extracting the
$n$-point NLSM amplitude -- the field-theory limit (\ref{ftlimX}) of
$Z_{1,\rho,n}(\tau)$ is enough to obtain the leading order of
(\ref{eqn:abelianZ}) in $\ap$. Moreover, one can identify the above
sine-functions with the string-theory momentum kernel 
\cite{momentumKernel}, defined recursively via \cite{abZtheory}
\beq
{\cal S}_{\alpha'}[A,j \, | \, B,j,C]_i =
\sin(\pi \alpha' k_{iB}\cdot k_{j}) \,{\cal S}_{\alpha'}[A \, | \, B,C]_i \co {\cal S}_{\alpha'}[\emptyset \, | \, \emptyset]_i \equiv 1  \ ,
\label{semiab59}
\eeq
with the same notation as seen in its field-theory counterpart
(\ref{momk}). More precisely, (\ref{eqn:abelianZ}) can be rewritten
in terms of its diagonal elements at rescaled value\footnote{The
rescaling stems from the present choice to incorporate the relative
monodromy phase between the two color factors in the brackets
$[t^i,t^j]_\ap$ of (\ref{eqn:commRelns}) via
$e^{i\pi \ap s_{ij}/2} t^i t^j - e^{-i\pi \ap s_{ij}/2}  t^j t^i$
instead of the more conventional representation
$ t^i t^j - e^{-i\pi \ap s_{ij}}  t^j t^i$ underlying
\cite{goodCDOpenString} and the original literature on
monodromy relations \cite{BjerrumBohr:2009rd, Stieberger:2009hq}.}
$\alpha'\rightarrow \alpha'/2$
\beq
Z_{\times}(\tau_n) = (2i)^{n-2} \sum_{\rho \in S_{n-2}} {\cal S}_{\alpha'/2}[ \rho(23\ldots n{-}1) \,
| \, \rho(23\ldots n{-}1) 
]_1 \, Z_{1,\rho(2,3,\ldots,n{-}1),n}(\tau_n) \ .
\label{eqn:preKLTMystery}
\eeq


\subsubsection{Deriving the BCJ numerators of the NLSM}

Here we can resolve a mystery first identified in
ref.~\cite{ckNLSM}, and made acute in ref.~\cite{abZtheory}.  In the
former reference it was shown that color-kinematic satisfying
numerators can be written down for the NLSM as some sum over
permuted entries of the momentum kernel matrix (\ref{momk}).  For theories compatable
with the color-kinematics duality, there exists representations of the amplitudes,
proven at tree-level, conjecturally to all loop order, where the kinematic weights of 
cubic graphs obey the same algebraic relationships as the color weights.
This means, order by order, a finite set of boundary graphs to all multiplicity, termed {\it master graphs}, 
encodes all the information of the amplitude,
and algebraic relations propagate this information to the remaining bulk-graphs.  For adjoint theories,
the algebraic relations are anti-symmetry around vertex flips (mirroring the behavior of adjoint structure constants), as well as Jacobi identities around
all edges.    Jacobi identites always express one graph's weight in terms of two other weights. 
In ref.~\cite{abZtheory} it was realized that in fact one needed only the diagonal elements of the
KLT matrix to construct the master numerators. The reason can be
understood by recalling the emergence of NLSM amplitudes from abelian
$Z$-theory \cite{abZtheory},
\beq
A^{\te{NLSM}}(\tau_n) = \lim_{\ap \rightarrow 0} \ap^{2-n} Z_{\times}(\tau_n) \ ,
\label{NLSM2}
\eeq
see (\ref{NLSM}), and inserting the $\ap\to0$ limits of the two
constituents in (\ref{eqn:preKLTMystery}), namely (\ref{ftlimX}) and
\begin{align}
 {\cal S}_{\alpha'/2}[ \rho(2\ldots n{-}1) 
|  \rho(2\ldots n{-}1) 
]_1 &= \left(\frac{ \pi \alpha' }{2} \right)^{n-2} \! \!  S[ \rho(2\ldots n{-}1) 
| \rho(2\ldots n{-}1) ]_1+ {\cal O}(\ap{}^{n}) \, .  \label{ftlim} 
\end{align}
I.e.\ the string-theory KLT matrix (\ref{semiab59}) and the doubly-ordered $Z$-amplitudes 
(\ref{sig04}) limit to the field-theory KLT matrix (\ref{momk}) and the
doubly-stripped bi-adjoint scalar amplitude $m[\cdot | \cdot ]$
\cite{Cachazo:2013iea}, respectively. We therefore obtain the compact formula
for NLSM master numerators proposed in \cite{abZtheory} 
\beq
A^{\te{NLSM}}(\tau_n) = (\pi i)^{n-2} \! \! \sum_{\rho \in S_{n-2}} \! \! S[ \rho(23\ldots n{-}1) \,
| \, \rho(23\ldots n{-}1) 
]_1 \, m[1,\rho(2,\ldots,n{-}1),n | \tau_n] \ ,
\label{semiab61}
\eeq
from the field-theory limit of (\ref{eqn:preKLTMystery}). As firstly
exploited implicitly in \cite{Mafra:2011kj}, color-kinematic
satisfying master numerators enter the full amplitude through a sum
over their product with the doubly-stripped partial amplitudes
$m[\cdot | \cdot ]$ of the bi-adjoint scalar theory. Hence, the role
of the diagonal entries $S[ \rho(23\ldots n{-}1) \,| \,
\rho(23\ldots n{-}1) ]_1$ in (\ref{semiab61}) identifies them as the
master numerators of the NLSM \cite{abZtheory}.

We would be remiss if we did not refer to a remarkable recent result due to Cheung and Shen~\cite{Cheung:2016prv}.  There an explicit cubic action was found for the NLSM which indeed generates
exactly these color-dual kinematic numerators from application of naive Feynman rules.

\subsection{Examples of semi-abelian $Z$-amplitudes}
\label{sect34}

In this section, we extract the ${\cal W}_\ap$-matrices in (\ref{structural}) from the
semi-abelian CP-dressed $Z$-amplitudes in their simplified representation (\ref{CPdressed})
when a small number $r=0,1,\ldots,5$ of CP factors is left
non-abelian.

\subsubsection{$r\le2$ non-abelian generators}

Given the cyclic equivalence of integration domains $\Sigma(12\ldots r) \rightarrow \Sigma(2\ldots r1)$, we need a minimum of $r=3$ non-trivial CP generators in (\ref{eqn:sabZ}) to deviate from the abelian disk integrals $Z_{\times}(\ldots)$: This can be immediately seen from the rearrangements of the integration region (\ref{semishuffle}) following from the definition (\ref{semiab02}) of $Z_1(\tau_n)$ and $Z_{12}(\tau_n)$,
\begin{align}
Z_{1}(\tau_n) 
=Z_{12}(\tau_n) &=  \sum_{\myatop{\sigma(2,3,\ldots,n) }{\in 2\shuffle 3 \shuffle \ldots \shuffle n}} Z_{1,\sigma(2,3,\ldots, n)}(\tau_n) 
= \sum_{\sigma \in S_{n-1}} Z_{1,\sigma(2,3,\ldots, n)}(\tau_n) 
= Z_{\times}(\tau_n) \, .
\label{semiab03}
\end{align}
Equivalently,
one can check (\ref{semiab03}) by comparing the sine-functions
in the trace (and its permutations in $2,3,\ldots,n{-}1$)
\begin{align}
&\te{Tr}( [ \ldots [[ t^1 , t^{2} ]_{\ap}  , t^{3}]_{\ap} , \ldots , t^{n-1}]_{\ap} t^n ) \, \big|_{t^{2},\ldots  ,t^{n-1} = {\mathbf 1}}  = (2i)^{n-2} \te{Tr}( t^1 t^n) \prod_{k=2}^{n-1} \sin(x_{12\ldots k-1,k})  \ ,
\label{semiab63}
\end{align}
with (\ref{eqn:abelianZ}) after stripping off the trace
$ \te{Tr}( t^1 t^n)$ of the leftover non-abelian generators.
Hence, the non-trivial semi-abelian disk integrals which are
different from their abelian counterparts involve at least
$r\geq 3$ non-abelian generators.


\subsubsection{$r=3$ non-abelian generators}

For three non-trivial CP generators at positions $i,j$ and $n$
and all other generators abelian, $t^{\ell\neq i,j,n}\rightarrow {\mathbf 1}$,
the CP-dressed $Z$-amplitudes (\ref{CPdressed}) boil down
to traces of the form
\begin{align}
&\te{Tr}( [ \ldots [ \ldots[ \ldots [ t^1 , t^{2} ]_{\ap}  ,\ldots , t^{i}]_{\ap} , \ldots , t^{j}]_{\ap}, \ldots ,t^{n-1}]_{\ap} t^n )  \label{semiab66} \\
&\ \ \ \ \ \rightarrow  (2i)^{n-3}  \prod_{\myatop{k=2 }{k\neq j}}^{n-1} \sin(x_{12\ldots k-1,k})  \te{Tr}(e^{i x_{12\ldots j-1,j} } t^i t^j t^n  - e^{-i x_{12\ldots j-1,j} }  t^j t^i t^n) \ ,
\notag
\end{align}
where the coefficients of $\te{Tr}( t^{\Sigma(i)} t^{\Sigma(j)} t^n)$
in ${\cal Z}(\tau)$ determines the semi-abelian integrals
$Z_{\Sigma(ij)n}(\tau)$. Since the latter are known to be
real, we will only be interested in the real part of (\ref{semiab66}), e.g.
\beq
\Re \Big[ (2i)^{n-3}  \prod_{\myatop{k=2 }{k\neq j}}^{n-1} \sin(x_{12\ldots k-1,k})  e^{i x_{12\ldots j-1,j} } \Big] =  (2i)^{n-3}  \prod_{\myatop{k=2}{k\neq j}}^{n-1} \sin(x_{12\ldots k-1,k}) \left\{
\begin{array}{ll}
i \sin(x_{12\ldots j-1,j}) &: \ n \ \te{even}
\\
\cos(x_{12\ldots j-1,j}) &: \ n \ \te{odd}
\end{array}
\right.
\label{semiab67} 
\eeq
along with $\te{Tr}( t^{i} t^{j} t^n)$. Note that $\cos(x_{12\ldots j-1,j})$ enters with a different sign when considering
$\te{Tr}( t^{j} t^{i} t^n)$ instead of $\te{Tr}( t^{i} t^{j} t^n)$.
As such we arrive at the overall result
\begin{align}
{\cal W}_\ap(\Sigma(ij)n\,|\,2\ldots i\ldots j\ldots n{-}1)  &= (2i)^{n-3} \prod_{\myatop{k=2}{k\neq j}}^{n-1} \sin(x_{12\ldots k-1,k})  \label{semiab68}  \\
&\! \! \! \! \! \! \! \! \! \! \! \! \! \! \! \! \! \! \! \! \! \! \! \! \! \! \! \! \!  \times \left\{
\begin{array}{ll}
i \sin (x_{12\ldots j-1,j}) &: \ n \ \te{even} \\
\te{sgn}(\Sigma(ij) | ij) \cos (x_{12\ldots j-1,j}) &: \ n \ \te{odd} 
\end{array}
\right.
\notag
\end{align}
for the ${\cal W}_\ap$-matrix in (\ref{structural}), where
$\te{sgn}(ij | ij)=1$ and $\te{sgn}(ji | ij)=-1$. For even
multiplicity $n=2m$, one recovers half the result ${\cal W}_\ap(ij
\,|\, 23\ldots 2m{-}1)= (2i)^{n-2}\prod_{k=2 }^{n-1}
\sin(x_{12\ldots k-1,k})$ known from two non-abelian CP factors
$t^i$ and $t^j$. Hence, the semi-abelian disk integrals for three
non-abelian CP factors and even $n$ are again captured by their abelian
counterparts\footnote{An alternative argument can be derived from
reflection symmetry $Z_{123\ldots
n}(\tau_{n})=(-1)^nZ_{1,n\ldots32}(\tau_{n})$.
},
\beq
Z_{1ij}(\tau_{2m})  = \frac{1}{2} \, Z_{\times}(\tau_{2m}) \ .
\label{new3g}
\eeq
The first novel expression for a semi-abelian integral (i.e.\ different from $Z_{\times}(\tau_n)$) can be found at five points with three non-abelian CP factors, where (\ref{semiab68}) implies that
\begin{align}
Z_{345}(\tau_5) &= 4 \big[ \sin(x_{1,2}) \sin(x_{12,4}) \cos(x_{124,3}) Z_{12435}(\tau_5) +  \sin(x_{1,4}) \sin(x_{14,2}) \cos(x_{124,3}) Z_{14235}(\tau_5)
 \notag \\
& \! \! \! \! \! \! \! \! \! \! \! \! \! \! \! \! \! \! \! \!  
+  \sin(x_{1,4}) \sin(x_{134,2}) \cos(x_{14,3}) Z_{14325}(\tau_5) 
-  \sin(x_{1,2}) \sin(x_{12,3}) \cos(x_{123,4}) Z_{12345}(\tau_5) 
   \label{semiab70}  \\
& \! \! \! \! \! \! \! \! \! \! \! \! \! \! \! \! \! \! \! \!  
-  \sin(x_{1,3}) \sin(x_{13,2}) \cos(x_{123,4}) Z_{13245}(\tau_5)
-  \sin(x_{1,3}) \sin(x_{134,2}) \cos(x_{13,4}) Z_{13425}(\tau_5)  \big]
 \ .
\notag
\end{align}
The two sine-factors in each term signal the leading low-energy
order $\ap^2$ and lead to the $\ap$-expansions
\begin{align}
Z_{345}(1,2,3,4,5) &= 
(\pi \ap)^2  \left(1 - \frac{s_{51}+s_{12} }{s_{34} }  - \frac{s_{23}+s_{12} }{s_{45} } \right)
+
\frac{(\pi \ap)^4}{12} \, \Big( 2s_{12} s_{23}
          + 2s_{12}^2
          + 3s_{51}s_{23}
           \notag \\
          &+ 2s_{51}s_{12}
          + s_{45}s_{23}
          - s_{45}s_{12}
          - 2s_{45}s_{51}
          - 2s_{34}s_{23}
          - s_{34}s_{12}
          + s_{34}s_{51}
          + 2 s_{34}s_{45}  \notag \\
          &- \frac{s_{12}^3+2s_{51}s_{12}^2 +2s_{51}^2s_{12} +s_{51}^3}{s_{34}} 
       - \frac{s_{12}^3+2s_{23}s_{12}^2 +2s_{23}^2s_{12} +s_{23}^3}{s_{45}} \Big) + {\cal O}(\ap^5) \notag 
\\
Z_{235}(1,2,3,4,5) &=
(\pi \ap)^2  \left(1  - \frac{ s_{45}+s_{51} }{s_{23}} \right)  +
 \frac{(\pi \ap)^4}{12} \, \Big(
           2s_{51}^2 - 2s_{51}s_{12}
          + s_{45}s_{12}
          + 6s_{45}s_{51}
          + 2s_{45}^2 \notag \\
          &
          + 3s_{34}s_{12}
          + s_{34}s_{51}
          - 2s_{34}s_{45}
          + 2s_{23}s_{12}
          - 3s_{23}s_{51}
          - 3s_{23}s_{45}
          + 2s_{23}s_{34}
          + 2s_{23}^2 \notag \\
          &
            - \frac{ s_{51}^3
          + 2s_{45}s_{51}^2
          + 2s_{45}^2s_{51}
          + s_{45}^3 }{ s_{23}}
          \Big) + {\cal O}(\ap^5)   \label{semiab05} 
          \end{align}   
after appropriate relabelling in the second case. The non-abelian $Z$-amplitudes on the right hand side of (\ref{semiab70}) can for instance be evaluated through the Berends--Giele techniques of \cite{nonabZ}.
Seven-point examples of the low-energy
limits at the order of $(\pi \ap)^4$ to be found in (\ref{freddy7})
and (\ref{freddy7b}) can be easily arrived at by inserting
(\ref{semiab68}) into (\ref{structural}).

As will be detailed in section \ref{sect4}, the low-energy limits $\sim (\pi \ap)^2$ of (\ref{semiab05}) tie in with the expressions in section 2.3 of \cite{Cachazo:2016njl},
\begin{align}
A_5(1^\phi,2^\phi,3^\phi,4^\Sigma,5^\Sigma) &= \frac{ s_{34}+s_{45} }{s_{12}} +  \frac{ s_{15}+s_{45} }{s_{23}} - 1
\notag
\\
A_5(1^\phi,2^\phi,3^\Sigma,4^\phi,5^\Sigma) &= \frac{ s_{34}+s_{45} }{s_{12}}  - 1  \ ,
\label{1604b}
\end{align}
which describe the doubly-ordered five-point amplitudes involving
two NLSM pions $\Sigma$ and three bi-colored $\phi^3$ scalars.


\subsubsection{$r=4$ and $r=5$ non-abelian generators}

The $r=4$ analogue of the trace (\ref{semiab66}) with
$t^{\ell\neq p,q,r,n}\rightarrow {\mathbf 1}$ is given by
\begin{align}
&\te{Tr}( [\ldots [ \ldots [ \ldots[ \ldots [ t^1 , t^{2} ]_{\ap}  ,\ldots , t^{p}]_{\ap} , \ldots , t^{q}]_{\ap}, \ldots ,t^{r}]_{\ap},\ldots ,t^{n-1}]_{\ap} t^n ) \notag \\
&\ \ \ \ \ \rightarrow  (2i)^{n-4}  \prod_{\myatop{k=2}{k\neq q,r}}^{n-1} \sin(x_{12\ldots k-1,k})  \te{Tr}\Big( e^{i x_{12\ldots r-1,r} } (e^{i x_{12\ldots q-1,q} }t^p t^q t^r  t^n  
-e^{-i x_{12\ldots q-1,q} } t^q t^p t^r  t^n)\notag \\
& \ \ \ \ \ \ \ \ \ \ \ \ \ \ \ \ \ \  \ \ \ \ \ \ \ \ \  \ \ \ \ \ \ \ \ \ 
- e^{-i x_{12\ldots r-1,r} }(e^{i x_{12\ldots q-1,q} } t^r t^p t^q   t^n
-e^{-i x_{12\ldots q-1,q} } t^r t^q t^p  t^n) \Big)
\notag \\
&\ \ \ \ \ =(2i)^{n-4}  \prod_{\myatop{k=2}{k\neq q,r}}^{n-1} \sin(x_{12\ldots k-1,k})  \te{Tr}\Big(
\cos ( x_{12\ldots q-1,q}) \cos(x_{12\ldots r-1,r})   [[t^p,t^q], t^r]  t^n  \label{semiab71} \\
& \ \ \ \ \ \ \ \ \ \ \ \ \ \ \ \ \ \  \ \ \ \ \ \ \ \ \  \ \ \ \ \ \ \ \ \  -\sin(x_{12\ldots q-1,q}) \sin(x_{12\ldots r-1,r}) \{ \{ t^p,t^q\}, t^r\} t^n  \notag \\
& \ \ \ \ \ \ \ \ \ \ \ \ \ \ \ \ \ \  \ \ \ \ \ \ \ \ \  \ \ \ \ \ \ \ \ \ + i\cos(x_{12\ldots q-1,q}) \sin(x_{12\ldots r-1,r})\{[t^p,t^q], t^r\} t^n  \notag \\
&  \ \ \ \ \ \ \ \ \ \ \ \ \ \ \ \ \ \  \ \ \ \ \ \ \ \ \  \ \ \ \ \ \ \ \ \  +i\sin(x_{12\ldots q-1,q}) \cos(x_{12\ldots r-1,r})   [ \{ t^p,t^q\}, t^r] t^n
\Big) \ . \notag
\end{align}
Selecting the real part of (\ref{semiab71}) amounts to constraining the
number of brackets accompanied by a sine-function, such that
\begin{align}
&{\cal W}_\ap(\Sigma(pqr)n \,|\, 23\ldots p\ldots q\ldots r\ldots n{-}1)  = (2i)^{n-4} \prod_{\myatop{k=2}{k\neq q,r}}^{n-1} \sin(x_{12\ldots k-1,k})  \label{semiab72}  \\
&\ \ \ \ \ \times \left\{
\begin{array}{ll}
\cos(x_{12\ldots q-1,q}) \cos(x_{12\ldots r-1,r}) \te{Tr}( [[t^p,t^q], t^r] t^n) \\
 -\sin(x_{12\ldots q-1,q}) \sin(x_{12\ldots r-1,r}) \te{Tr}(\{ \{ t^p,t^q\}, t^r\} t^n)
\big|_{\te{Tr}( t^{\Sigma(p)} t^{\Sigma(q)} t^{\Sigma(r)} t^n)} &: \ n \ \te{even} \\ \\
i\cos(x_{12\ldots q-1,q}) \sin(x_{12\ldots r-1,r}) \te{Tr}( \{[t^p,t^q], t^r\} t^n) \\
+i\sin(x_{12\ldots q-1,q}) \cos(x_{12\ldots r-1,r}) \te{Tr}( [ \{ t^p,t^q\}, t^r] t^n)
\big|_{\te{Tr}( t^{\Sigma(p)} t^{\Sigma(q)} t^{\Sigma(r)} t^n)}
 &: \ n \ \te{odd} 
\end{array}
\right. \ . \notag
\end{align}
The notation $(Y)|_{\te{Tr(X)}}$ instructs to select from the
expression $Y$ the coefficients of the CP trace $\te{Tr}(X)$.  The
obvious $r=5$ counterpart of (\ref{semiab71}) with $t^{\ell\neq
p,q,r,s,n}\rightarrow {\mathbf 1}$ yields
\begin{align}
&{\cal W}_\ap(\Sigma(pqrs)n \, | \, 23\ldots p\ldots q\ldots r\ldots s \ldots n{-}1)  = (2i)^{n-5} \prod_{\myatop{k=2}{k\neq q,r,s}}^{n-1} \sin(x_{12\ldots k-1,k})   \label{semiab73} \\
& \ \times \left\{
\begin{array}{ll}
\te{Tr}(\cos(x_{12\ldots q-1,q}) \cos(x_{12\ldots r-1,r}) \cos(x_{12\ldots s-1,s}) [ [ [t^p,t^q], t^r] , t^s] t^n\\
-\cos(x_{12\ldots q-1,q}) \sin(x_{12\ldots r-1,r}) \sin(x_{12\ldots s-1,s}) \{ \{ [t^p,t^q], t^r\} , t^s\} t^n
\ \ \ \ \ \ \ \ \ \ \ \ \ \ \ \ \ \ \  : \ n \ \te{odd} \\
-\sin(x_{12\ldots q-1,q}) \cos(x_{12\ldots r-1,r}) \sin(x_{12\ldots s-1,s}) \{ [ \{ t^p,t^q\}, t^r] , t^s\} t^n \\
-\sin(x_{12\ldots q-1,q}) \sin(x_{12\ldots r-1,r}) \cos(x_{12\ldots s-1,s}) [ \{ \{ t^p,t^q\}, t^r\} , t^s] t^n) \big|_{\te{Tr}( t^{\Sigma(p)} t^{\Sigma(q)} t^{\Sigma(r)}t^{\Sigma(s)} t^n)}
  \\ \\
\te{Tr}(-i\sin(x_{12\ldots q-1,q}) \sin(x_{12\ldots r-1,r}) \sin(x_{12\ldots s-1,s}) \{ \{ \{ t^p,t^q \}, t^r\} , t^s \} t^n \\
+i\sin(x_{12\ldots q-1,q}) \cos(x_{12\ldots r-1,r}) \cos(x_{12\ldots s-1,s}) [[\{t^p,t^q\}, t^r] , t^s] t^n
\ \ \ \ \ \ \ \ \ \ \ \ \ \ \ \ \ \ \  : \ n \ \te{even}    \\
+i\cos(x_{12\ldots q-1,q}) \sin(x_{12\ldots r-1,r}) \cos(x_{12\ldots s-1,s}) [ \{ [  t^p,t^q], t^r\} , t^s] t^n \\
+i\cos(x_{12\ldots q-1,q}) \cos(x_{12\ldots r-1,r}) \sin(x_{12\ldots s-1,s}) \{ [ [ t^p,t^q], t^r] , t^s\} t^n)
\big|_{\te{Tr}( t^{\Sigma(p)} t^{\Sigma(q)} t^{\Sigma(r)}t^{\Sigma(s)} t^n)}
\end{array}
\right. \notag
\end{align}
Examples for five- and six-point low-energy limits with $r=4$
can be found in (\ref{semiab06}) as well as
(\ref{freddy}) to (\ref{endfreddy}), respectively. Moreover, the
leading $\ap$-orders of six-point integrals with $r=5$ are
displayed in (\ref{65a}) to (\ref{65b}).


\subsection{General form of the semi-abelian ${\cal W}_{\alpha'}$-matrix}
\label{genform}

To conjecture a form of the ${\cal W}_{\alpha'}$-matrix in (\ref{structural}) for an arbitrary number of non-trivial CP particles it is helpful to introduce a unifying notation relying on a set of binary vectors,
\begin{align}
\label{eq:binAB}
{\rm Bin}(a,b) &\equiv \{ v\in (\{0,1\})^a ~ {\rm s.t.}~ |v|^2 ~{\rm odd} \iff b ~{\rm odd}  \} \, . 
\end{align}
 ${\rm Bin}(a,b)$ is the set of binary vectors in an $a$-dimensional space whose magnitude squared is odd if and only if $b$ is odd, e.g.
\begin{align}
{\rm Bin}(3,1) ={\rm Bin}(3,3) &=  \left\{ (1,0,0), \, (0,1,0), \, (0,0,1), \, (1,1,1) \right\}  \\
{\rm Bin}(3,0) ={\rm Bin}(3,2) &=  \left\{ (1,1,0), \, (1,0,1), \, (0,1,1), \, (0,0,0) \right\} \, .
\end{align}
Using the binary-vector notation, the above derivations are consistent with a ${\cal W}_{\alpha'}$ given as:
\begin{multline}
{\cal W}_{\alpha'}(\Sigma(p_1 p_2 \cdots p_{r-1} )n  \, |\, 23\ldots p_1 \ldots p_2 \ldots\ldots p_{r-1} \ldots ( n{-}1) )=(2 i)^{n-r} \!\!\!\!\!\! \prod^{n-1}_{ {\myatop{k=2}{k \neq   p_2,\cdots ,p_{r-1}}} }  \!\!\!\!\!\!  \sin\left(x_{12\ldots (k-1) ,\, k}\right) \\
\times \!\!\!\! \sum_{v \in {\rm Bin}(r-2,n)} \!\!\!\!  \left.{\rm Tr}\left( \left[\left[\cdots \left[ \left[p_1,\underline{p_2 }\,\right]_{v_1}, \underline{p_3}\,\right]_{v_2}
\cdots \right]_{v_{r-3}},\, \underline{p_{r-1}}\, \right]_{v_{r-2}}  t^n \right)\right|_{{\rm Tr}\left(t^{\Sigma(p_1)} \cdots t^{\Sigma(p_{r-1})}\,t^n \right)} \!\!\!\!\!\! , 
\label{binv}
\end{multline}
where without loss of generality we take the first leg to be CP-abelian, leg $n$ to be CP-non-abelian, and the second entry of ${\cal W}_\ap$ to be canonically ordered. The binary commutators (whose trigonometric dressing singles out the underlined entry $p_b$) are defined as follows,
\begin{align}
[p_a, \underline{p_b}\,]_{0} &\equiv  \left( t^{p_a} t^{p_b}  + t^{p_b} t^{p_a} \right) \times i \sin(x_{12 \ldots  (p_b-1),\, p_b})  \label{eqn:trigCom} \\
[p_a, \underline{p_b}\,]_{1} &\equiv  \left(  t^{p_a} t^{p_b}   -  t^{p_b} t^{p_a}  \right) \times \cos(x_{12 \ldots  (p_b-1),\, p_b})   \notag \,.
\end{align}


\subsection{Structure of the low-energy expansion}
\label{sect37}

In this subsection, we describe the implications of the
representation (\ref{structural}) of semi-abelian disk integrals
(\ref{semiab02}) for the structure of their low-energy expansion. As
emphasized in (\ref{semiab57}), each sine-factor descending from the
$\ap$-weighted brackets in (\ref{CPdressed}) contributes an overall
factor of $\pi \ap$. For $r$ non-abelian CP factors and $n$ external
legs, tracking the commutators $[\cdot,\cdot]_{\ap}$ with an identity
matrix in one of their entries amounts to the lower bound $(\pi \ap)^{n-r}$ on the
leading low-energy order, see the examples in the previous section.
Moreover, depending on $(2i)^{n-r} $ being real or imaginary,
another sine factor with low-energy order $\pi \ap$ arises from the $\ap$-weighted traces,
leading to the refined lower bound
\beq
Z_{\Sigma(12\ldots r)}(\tau_n) = \left\{ \begin{array}{cl}
{\cal O}(\pi \ap^{n-r}) &: \ n-r \ \te{even} \ , \ \ r\geq 2 \\
{\cal O}(\pi \ap^{n-r+1}) &: \ n-r \ \te{odd} \ , \ \ \ r\geq 3
\end{array} \right. \ .
\label{lower}
\eeq
For small $n$ and $r$, (\ref{lower}) implies the following leading
low-energy contributions for $Z_{\Sigma(12\ldots r)}(\tau_n)$,
\beq
\begin{array}{|c||c|c|c|c  |c|c|c  |}
\hline  
\ n \ \ & \ r\leq 2 \ \ & \ r=3 \ \ & \ r=4 \ \ & \ r=5 \ \
& \ r=6 \ \ & \ r=7 \ \ & \ r=8 \ \ \\\hline \hline
4 &\alpha'^2 \zeta_2 &\alpha'^2 \zeta_2 &1 &\times
&\times &\times &\times \\  \hline
5 &0 &\alpha'^2 \zeta_2 &\alpha'^2 \zeta_2 &1
&\times &\times &\times \\  \hline
6 &\alpha'^4 \zeta_4 &\alpha'^4 \zeta_4 &\alpha'^2 \zeta_2  &\alpha'^2 \zeta_2
&1 &\times &\times \\  \hline
7 &0 &\alpha'^4 \zeta_4 &\alpha'^4 \zeta_4 &\alpha'^2 \zeta_2
&\alpha'^2 \zeta_2 &1 &\times \\  \hline
8 &\alpha'^6 \zeta_6 &\alpha'^6 \zeta_6 &\alpha'^4 \zeta_4  &\alpha'^4 \zeta_4
&\alpha'^2 \zeta_2 &\alpha'^2 \zeta_2 &1 \\  \hline
9 &0 &\alpha'^6 \zeta_6 &\alpha'^6 \zeta_6 &\alpha'^4 \zeta_4
&\alpha'^4 \zeta_4 &\alpha'^2 \zeta_2 &\alpha'^2 \zeta_2 \\\hline
\end{array} \ ,
\label{tableorder}
\eeq
unless special choices of $\Sigma$ and $\tau_n$ lead to additional
cancellations (see appendix \ref{app:expl}, in particular
(\ref{except}), (\ref{except2}) and (\ref{7shock}) for examples). Smaller values of
$r$ than admitted in (\ref{lower}) are already accounted for by
(\ref{semiab03}) and yield the abelian
integrals $Z_{\times}(\tau_n) = {\cal O}(\ap^{n-2})$ in
(\ref{eqn:abelianZ}). 

The even powers of $\pi$ in the leading low-energy orders
(\ref{lower}) can be obtained from rational multiples of Riemann
zeta values $\zeta_{2k}$ in the $\ap$-expansion of completely
ordered disk integrals $Z_\sigma(\tau_n)$ in (\ref{sig04}), e.g.
\beq
\zeta_2 = {\pi^2 \over 6}\,, \ \ \ \ \ 
\zeta_4 = {\pi^4 \over 90}\,, \ \ \ \ \ 
\zeta_6 = {\pi^6 \over 945} \,, \ \ \ \ \  \ldots \ \ \ \
\zeta_{2k} = (-1)^{k-1} { (2\pi)^{2k} B_{2k} \over 2 (2k)!}\,,
\label{evenzeta}
\eeq
with $B_{2k}$ denoting the Bernoulli numbers. When naively
assembling their semi-abelian counterparts (\ref{semiab02}) from
combinations of $Z_\sigma(\tau_n)$ via rearrangements (\ref{semishuffle}) of the
integration domain, the leading order of $(\pi \ap)^{2k}$ reflects
cancellations among the contributing $Z_\sigma(\tau_n)$ at all lower
orders $\ap^{\leq 2k-1}$, see section 4.3 of \cite{abZtheory}.

As a major advantage of the trigonometric representation
(\ref{structural}) of the semi-abelian disk integrals, their leading
low-energy order (\ref{lower}) is determined from the field-theory
limit (\ref{ftlimX}) of completely ordered disk integrals
$Z_\sigma(\tau)$. More generally, the $n{-}r$ and $n{-}r{+}1$ overall powers
of $\ap$ in the ${\cal W}_\ap$-matrix for even and odd numbers of
abelian CP factors, respectively, reduce the required order of
$\ap$ in the low-energy expansion of $Z_\sigma(\tau)$ by the same
amount when assembling the $\ap$-expansion of semi-abelian disk
integrals.

The $\ap$-expansion of $Z_\sigma(\tau)$\footnote{
Both the initial studies of $\ap$-expansions beyond four points
\cite{Medina:2002nk, Barreiro:2005hv, Oprisa:2005wu,
Stieberger:2006te} and powerful recent results on five-point
integrals \cite{Boels:2013jua, Puhlfuerst:2015gta} benefit from the
connection with (multiple Gaussian) hypergeometric functions.
} involves multiple zeta values (MZVs)
\beq
\zeta_{n_1,n_2,\ldots, n_r} \equiv \sum_{0<k_1<k_2<\ldots <k_r}^{\infty} k_1^{-n_1} k_2^{-n_2} \ldots k_r^{-n_r} 
\ , \ \ \ \ n_r \geq 2
\label{mzvdef}
\eeq
in a uniformly transcendental pattern, i.e.\ the order of $\ap^{w}$ is accompanied by MZVs of transcendental weight $w=n_1+n_2+\ldots+n_r$ \cite{Aomoto, Terasoma, Brown:2009qja, Stieberger:2009rr, Mafra:2011nw, Schlotterer:2012ny}. Uniform transcendentality is particularly transparent from the recursive method of \cite{Broedel:2013aza} to obtain the $\ap$-expansion of $n$-point integrals from the Drinfeld associator\footnote{Also see \cite{Drummond:2013vz} for a connection with the pattern relating the appearance of MZVs $\zeta_{n_1,n_2,\ldots,n_r}$ of different depth $r$ in open-superstring amplitudes \cite{Schlotterer:2012ny}.} acting on their $(n{-}1)$-point counterparts. Extending an alternative all-multiplicity technique based on polylogarithm integration \cite{Zfunctions}, a Berends--Giele recursion for the $\ap$-expansion of disk integrals $Z_\sigma(\tau)$ was given in \cite{nonabZ} whose efficiency comes to maximal fruition at high multiplicity and fixed order in $\ap$. 

By (\ref{structural}), all these expansion-methods for $Z_\sigma(\tau)$ as well as the results available for download via \cite{wwwMZV, nonabZ} can be neatly imported to infer the $\ap$-dependence of semi-abelian disk integrals. The complete (conjectural) basis of MZVs over $\mathbb Q$ present in the $\ap$-expansion of $Z_\sigma(\tau)$ generically enters their semi-abelian counterparts, accompanied by an appropriate global prefactor of $(\pi \ap)^{2k}$ as determined by (\ref{lower}). The coefficient of each such basis MZV in the semi-abelian $Z$-amplitudes signals an independent effective higher-derivative interaction between NLSM pions and $\phi^3$ scalars


\section{NLSM coupled to bi-adjoint scalars in semi-abelian $Z$-theory}
\label{sect4}

\subsection{Summary and overview}

In this section, we identify the low-energy limits of semi-abelian $Z$-theory amplitudes (\ref{semiab02}) with doubly partial amplitudes in a scalar field theory. We recall that the tree-level S-matrices of the bi-adjoint $\phi^3$ theory and the NLSM emerge from the $(\ap\rightarrow 0)$-regime of completely ordered disk integrals (\ref{sig04}) and their abelianized contribution (\ref{eqn:abelianZ}), respectively. On these grounds, it is not at all surprising that the ``interpolating'' case of semi-abelian disk integrals incorporates couplings between NLSM pions and $\phi^3$ scalars.  

While the bi-colored $\phi^3$ scalars are taken to be charged under two gauge groups with generators $t^a \otimes T^b$, the CP matrix $t^a$ is absent in the color-dressing $T^b$ of NLSM pions. In any field theory with interaction vertices involving both species, the tree-level amplitudes of $r$ bi-colored scalars and $n{-}r$ pions admit a color-decomposition 
\begin{align}
M_{r,n}^{{\rm NLSM} + \phi^3} 
&= \sum_{\sigma \in S_{r-1}} \sum_{\tau \in S_{n-1}}
\Tr(t^1t^{\sigma(2)}t^{\sigma(3)}\cdots t^{\sigma(r)}) 
\Tr(T^1T^{\tau(2)}T^{\tau(3)}\cdots T^{\tau(n)})  \notag \\
& \ \ \ \ \ \ \ \ \ \ \ \times
A^{{\rm NLSM} + \phi^3}(1,\sigma(2,3,\ldots,r)  \, |\, 1,\tau(2,3,\ldots,n))
\label{FTLIM1}
\end{align}
modulo multi-traces in the $t^a$ due to the exchange of pions in the
internal propagators. As a main result of this work, it is apparent that,
for a suitable choice of NLSM$+ \phi^3$ couplings, the
doubly-partial single-trace amplitudes
$A^{{\rm NLSM} + \phi^3}(1,\sigma(2,\ldots,r))  \, |\, 1,\tau(2,\ldots,n))$
in (\ref{FTLIM1}) emerge from the low-energy limit of semi-abelian
$Z$-theory amplitudes (\ref{semiab02}),
\beq
A^{{\rm NLSM} + \phi^3} (1,2,\ldots,r\, |\, \tau(1,2,\ldots,n)) = \lim_{\ap \rightarrow 0} (\ap)^{-2\lceil \frac{ n-r}{2} \rceil} Z_{12\ldots r}(\tau(1,2,\ldots,n)) \ ,
\label{ft01}
\eeq
which is equivalent to (\ref{NLSMphi3}). The structure of a
multi-trace completion of (\ref{FTLIM1}) as well as its tentative
string-theory origin (see e.g.~ref.~\cite{Green:1995ga}) is left as an interesting open problem for the
future.

A specific set of couplings NLSM$+ \phi^3$ is singled out
by the coefficients of the Adler zeros in the tree-level amplitudes
of the NLSM \cite{Cachazo:2016njl}. In this reference, interactions
of the NLSM with $\phi^3$ scalars are inferred from a soft-limit
extension of the NLSM, and the resulting single-trace doubly-partial amplitudes
are represented in the CHY framework \cite{Cachazo:2013gna,
Cachazo:2013hca, Cachazo:2013iea}. For an even number of pions, we claim that the
tree amplitudes of the NLSM$+ \phi^3$ theory in
\cite{Cachazo:2016njl} match the low-energy limits
(\ref{ft01}) of semi-abelian disk integrals, see (\ref{semiab05}) and (\ref{1604b}) 
for five-point examples. For odd values of
$n{-}r$, however, the NLSM$+ \phi^3$ amplitudes of
\cite{Cachazo:2016njl} vanish and do not admit any non-trivial
comparison with the leading $\ap$-order of semi-abelian $Z$-theory.

From the incarnation of relevant double-copy structures in the CHY formalism~\cite{Cachazo:2016njl},
the NLSM$+ \phi^3$ theory under investigation is closely
related to DBIVA coupled to sYM by dualizing 
the color-factors built
from the field-theory $T^a$ into kinematic factors of sYM.
The DBIVA $+$ sYM theory, in turn, appears in the low-energy
limit of string theory with abelian and non-abelian CP factors
in the tree amplitudes (\ref{openKLT}) and therefore supports the identification \eqref{ft01}.  

In the rest of this section, we discuss two implications of
(\ref{ft01}) for new representations of the tree-level S-matrix in
the NLSM$+\phi^3$ theory. 

\subsection{Amplitude relations: NLSM$+\phi^3$ versus pure $\phi^3$} 

The representation of semi-abelian disk integrals in (\ref{structural}) along with the low-energy limit of the ${\cal W}_{\ap}$-matrix therein reduces any doubly-partial amplitude of the coupled NLSM$+ \phi^3$ theory to those of pure $\phi^3$,
\beq
A^{{\rm NLSM} + \phi^3}( 1,2,\ldots ,r \, | \, \tau_n) =  \!\sum_{\rho \in S_{n-2}} \! \!  W( 12\ldots r \, | \, \rho(23\ldots n{-}1)) \,  m[1,\rho(23\ldots n{-}1),n \, | \, \tau_n] \ ,
\label{ft02}
\eeq
where $\tau_n \equiv \tau(1,2,\ldots,n)$. The entries of the $W$-matrix are polynomials in the Mandelstam invariants,
\beq
W( 12\ldots r \, | \, \rho(23\ldots n{-}1)) \equiv  \lim_{\ap \rightarrow 0} (\ap)^{-2\lceil \frac{ n-r}{2} \rceil}  {\cal W}_{\ap}( 12\ldots r \, | \, \rho(23\ldots n{-}1))  \ ,
\label{ft02a}
\eeq
which follow by replacing the trigonometric functions at the leading $\ap$-order of ${\cal W}_{\ap}$ via $\sin( x_{A,B}) \rightarrow \frac{i\pi}{2} k_A \cdot k_B $ and $\cos( x_{A,B}) \rightarrow 1$. In the simplest case of $r\leq 2$ bi-adjoint scalars, they coincide with the diagonal entries of the field-theory KLT matrix (\ref{momk}),
\begin{align}
W(\emptyset \, | \, 23\ldots n{-}1) &= 
W(p \, | \, 23\ldots n{-}1) = 
W(pq \, | \, 23\ldots n{-}1) \label{ft02b}\\
& = 
( i\pi)^{n-2} S[ 23\ldots n{-}1\, | \,  23\ldots n{-}1]_1
  = (i \pi)^{n-2}  \prod_{j=2}^{n-1} k_{12\ldots j-1} \cdot k_{j} \ ,
\notag
\end{align}
and do not depend on the legs $p,q$.
The resulting amplitude relations connecting the bi-adjoint scalar theory with its coupling to the NLSM read
\begin{align}
&A^{{\rm NLSM} + \phi^3}( \emptyset \, | \, \tau_n) = A^{{\rm NLSM} + \phi^3}( 1 \, | \, \tau_n) = A^{{\rm NLSM} + \phi^3}( 1,2 \, | \, \tau_n)  \label{ft02c} \\
& \ \ \ = (i \pi)^{n-2} 
\sum_{\rho \in S_{n-2}}  m[1,\rho(2,3,\ldots,n{-}1),n \,| \, \tau_n]  
 \prod_{j=2}^{n-1} (k_{1\rho(23\ldots j-1)} \cdot k_{\rho(j)}) \ ,
\notag
\end{align}
and their general form (\ref{ft02}) resembles recent relations \cite{Stieberger:2016lng, Nandan:2016pya, delaCruz:2016gnm, Schlotterer:2016cxa, Du:2016wkt} between Einstein--Yang--Mills amplitudes and those of pure Yang--Mills theory.
We shall next elaborate on the cases with $r\geq 3$ bi-colored
scalars and extract the field-theory limits (\ref{ft02a}) from the 
${\cal W}_{\alpha'}$ matrices  of the previous section via
$\sin(x_{12\ldots j-1,j}) \rightarrow \frac{i \pi}{2}
(k_{12\ldots j-1}\cdot k_{j})$ and $\cos(x_{12\ldots j-1,j})
\rightarrow 1$.

\subsubsection{$r=3,4$ bi-adjoint scalars}

With three or four bi-adjoint scalars at legs $n,p,q,r$, (\ref{semiab68}) and (\ref{semiab72}) imply
\beq
W(\Sigma(pq)n \, | \, 23\ldots p\ldots q\ldots n{-}1)  = (i\pi)^{n-3} \prod_{ \myatop{j=2}{j\neq q}}^{n-1} k_{12\ldots j-1} \cdot k_{j} \times \left\{
\begin{array}{ll}
\frac{i\pi}{2} \, k_{12\ldots q-1}\cdot k_{q} &: \ n \ \te{even} \\
\te{sgn}(\Sigma(pq) | pq) &: \ n \ \te{odd} 
\end{array}
\right.
\label{semiab68z} 
\eeq
with $\te{sgn}(pq | pq)=1$ and $\te{sgn}(qp | pq)=-1$
as well as
\begin{align}
&W(\Sigma(pqr)n \, | \, 23\ldots p\ldots q\ldots r\ldots n{-}1) = (i\pi)^{n-4} \prod_{\myatop{j=2}{j\neq q,r}}^{n-1} k_{12\ldots j-1}\cdot k_{j}  \label{semiab72z}  \\
&\ \ \ \ \ \times \left\{
\begin{array}{ll}
\te{Tr}(    [ [t^p ,t^q], t^r ] t^n)
\big|_{\te{Tr}( t^{\Sigma(p)} t^{\Sigma(q)} t^{\Sigma(r)} t^n)} &: \ n \ \te{even} \\ \\
\frac{i \pi }{2} \,(k_{12\ldots q-1}\cdot k_{q}) \te{Tr}(    [  \{ t^p ,t^q\}, t^r ] t^n)
\\
+\frac{i \pi}{2}\, (k_{12\ldots r-1}\cdot k_{r})   \te{Tr}(     \{ [t^p ,t^q], t^r \} t^n)
\big|_{\te{Tr}( t^{\Sigma(p)} t^{\Sigma(q)} t^{\Sigma(r)} t^n)}
 &: \ n \ \te{odd} 
\end{array}
\right. \notag
\end{align}
Note that cases with three bi-adjoint scalars at legs $p,q,n$ and even multiplicity simplify to
\beq
W( \Sigma(pq)n \, | \, 23\ldots p\ldots q\ldots n{-}1) \, \Big|_{n \ \te{even}} = \frac{1}{2}  \, W( \emptyset \, | \, 23\ldots n{-}1)
\label{semiab68y}
\eeq
by the first line of (\ref{semiab68z}).

\subsubsection{General form of the $W$-matrix in field theory}

In the binary-vector representation of the ${\cal W}_{\alpha'}$-matrix given in section \ref{genform}, words with large numbers of entries $v_j=1$ dominate the sum in (\ref{binv}). For even numbers $n{-}r$ of pions, only the word $v= (1,1,\ldots,1)$ contributes and yields the simple result
\begin{align}
&W(\Sigma(p_1 p_2 \cdots p_{r-1} )n  \, |\, 23\ldots p_1 \ldots p_2 \ldots\ldots p_{r-1} \ldots ( n{-}1) ) \, \big|_{n-r  \ {\rm even} }
=(i\pi)^{n-r} 
\label{genFTW1}  \\
& \ \ \ \  \times  \bigg( \! \! \! \! \! \prod^{n-1}_{  \myatop{j =2}{j \neq   p_2,\cdots ,p_{r-1}} }  \!\!\!\!\!\!  k_{12\ldots (j-1)}\cdot  k_j \bigg) \
{\rm Tr}( [[ \ldots [[ t^{p_1} , t^{p_2} ], t^{p_3} ] , \ldots ], t^{p_{r-1}} ] t^n     ) \, \big|_{  {\rm Tr}(  t^{\Sigma(p_1)} t^{\Sigma(p_2)} \ldots t^{\Sigma(p_{r-1})} t^n   )  } \notag
\end{align}
in terms of commutators. For odd values of $n{-}r$, on the other hand, the leading low-energy order of (\ref{binv}) stems from words with a single entry $v_{\ell}=0$ such that the trace in
\begin{align}
&W(\Sigma(p_1 p_2 \cdots p_{r-1} )n  \, |\, 23\ldots p_1 \ldots p_2 \ldots\ldots p_{r-1} \ldots ( n{-}1) ) \, \big|_{n-r  \ {\rm odd} }
 \notag \\
& \ \ \ \  = \frac{1}{2} \, (i\pi)^{n-r+1}  \bigg( \! \! \! \! \! \prod^{n-1}_{  \myatop{j =2}{j \neq   p_2,\cdots ,p_{r-1}} }  \!\!\!\!\!\!  k_{12\ldots (j-1)}\cdot  k_j \bigg) \
\sum_{\ell=2}^{r-1} (k_{12\ldots (p_\ell - 1)} \cdot k_{p_\ell}) 
\label{genFTW2}
\\
& \ \ \ \ \ \ \ \  \times 
{\rm Tr}( [[ \ldots [ \{ [ \ldots [[ t^{p_1} , t^{p_2} ], t^{p_3} ] , \ldots ], t^{p_\ell} \} , t^{p_{\ell+1}} ] , \ldots  ], t^{p_{r-1}} ] t^n     ) \, \big|_{  {\rm Tr}(  t^{\Sigma(p_1)} t^{\Sigma(p_2)} \ldots t^{\Sigma(p_{r-1})} t^n   )  }\notag
\end{align}
exhibits one anti-commutator operation $\{ \cdot , \cdot \}$ inside the nested commutators.


\subsection{Comparison with CHY integrands}

Recently the modern connected formalism of Cachazo, He and Yuan (CHY) \cite{Cachazo:2013gna, Cachazo:2013hca, Cachazo:2013iea} has given rise to all-multiplicity representations for NLSM amplitudes \cite{Cachazo:2014xea} and their (NLSM$+\phi^3$) extensions \cite{Cachazo:2016njl}. These CHY representations arise from integrals over the moduli space of punctured Riemann spheres, where the integrands depend on both the external data $\{t^{a_i},T^{b_i},k_i\}$ of the NLSM- or $\phi^3$ scalars and the punctures $z_i \in \mathbb C$ associated with the $i^{\rm th}$ leg. The punctures are constrained by the scattering equations
\beq
E_i \equiv \sum_{j\neq i}^n \frac{s_{ij}}{z_{ij}} = 0
\label{scatE}
\eeq
which mirror integration-by-parts relations in string theory and completely localize the integrals.

For any combination of the two species of scalars, the CHY integrands for NLSM$+ \phi^3$ amplitudes in \cite{Cachazo:2016njl} allow to factor out a universal piece, where $n$-particle Parke--Taylor factors $(z_{12} z_{23}\ldots z_{n,1})^{-1}$ are combined with traces $\Tr(T^{b_1} T^{b_2}\ldots T^{b_n})$ of the generators $T^{b_i}$ of the common gauge group. The other factor of the integrand depending on the number of pions and $\phi^3$ scalars is based on a matrix $A=A(\{k_i,z_i\})$ specified in \cite{Cachazo:2014xea}. For pure NLSM amplitudes, this non-universal piece of the CHY integrand is a reduced Pfaffian $(\te{Pf} \,'  A)^2$, where the prime refers to the deletion of two rows and columns each. For generic configurations of the two scalar species, the non-universal part of the integrand factorizes into an $r$-particle Parke--Taylor factor $(z_{12} z_{23}\ldots z_{r,1})^{-1}\Tr(t^{a_1} t^{a_2}\ldots t^{a_r})$ and a Pfaffian $(\te{Pf} \, A_{r{+}1,\ldots,n})^2$ referring to the external pion legs $r{+}1,\ldots,n$. Accordingly, the integrands vanish for an odd number $n{-}r$ of pions. 

In order to express the connected amplitudes in terms of doubly-partial amplitudes as done in (\ref{ft02}), any $z$-dependence from $(\te{Pf} \,'  A)^2$ and $(z_{12} z_{23}\ldots z_{r,1})^{-1}(\te{Pf} \, A_{r{+}1,\ldots,n})^2$ has to be reduced to Parke--Taylor factors -- $(z_{12} z_{23}\ldots z_{n,1})^{-1}$ and permutations in $1,2,\ldots,n$. The naive evaluation of the Pfaffians, however, involves more diverse functions of $z_j$ than captured by linear combinations of $\tau(z_{12} z_{23}\ldots z_{n,1})^{-1}$ with $\tau \in S_n$. The desired reduction to Parke--Taylor factors requires manifold applications of the scattering equations (\ref{scatE}) and can in principle be addressed through the algorithms of \cite{Cachazo:2015nwa, Cardona:2016gon}. Still, the complexity of these manipulations grows rapidly with the multiplicity and has therefore obstructed a compact Parke--Taylor representation of the non-universal integrands with more than four legs.

From the amplitude relations (\ref{ft02}) reducing the tree-level S-matrix of the NLSM$+ \phi^3$ theory to doubly-partial amplitudes, one can reverse-engineer a Parke--Taylor form of the underlying CHY integrands, valid on the support of the scattering equations (\ref{scatE}). The simple form for the $W$-matrix of NLSM amplitudes in (\ref{ft02b}) translates into the following representation of the connected integrand,
\begin{align}
(\te{Pf} 'A)^2 &= \sum_{\rho \in S_{n-2}} \frac{S[\rho(23\ldots n{-}1) | \rho(23\ldots n{-}1)]_1}{(1,\rho(2),\rho(3),\ldots,\rho(n{-}1),n) }  \ {\rm mod} \ E_i \notag \\
& = \sum_{\rho \in S_{n-2}} \frac{
  \prod_{j=2}^{n-1} (k_{1\rho(23\ldots j-1)} \cdot k_{\rho(j)})
 }{ (1,\rho(2),\rho(3),\ldots,\rho(n{-}1),n)} \ {\rm mod} \ E_i   \ ,
\label{chy1}
\end{align}
in terms of Parke--Taylor factors with
\beq
(1,2,3,\ldots,n{-}1,n) \equiv z_{12} z_{23} \ldots z_{n-1,n} z_{n,1} \ .
\label{chyPT}
\eeq
Similarly, (\ref{ft02}) along with the explicit $W$-matrices given in the previous section yields a simplified form of the connected integrands for mixed amplitudes $A^{{\rm NLSM} + \phi^3} (\Sigma(p_1 p_2\ldots p_{r-1})n\, |\, \ldots) $ with an even number of pions
\beq
\frac{(\te{Pf} \, A_{ \{ 12\ldots n{-}1 \}   \setminus \{ p_1,p_2,\ldots,p_{r-1} \}   })^2}{ (\Sigma(p_1),\Sigma(p_2), \ldots, \Sigma(p_{r-1}),n  )} = (i\pi)^{r-n} \sum_{\rho \in S_{n-2}} \frac{W ( \Sigma(p_1 p_2\ldots p_{r-1})n \, | \, \rho(23\ldots n{-}1))}{(1,\rho(2),\rho(3),\ldots ,\rho(n{-}1),n)} 
\ {\rm mod} \ E_i \ .
\label{chy2}
\eeq
Note that one can use the expression (\ref{genFTW1}) for the $W$-matrix, given the even values of $n{-}r$ in (\ref{chy2}).
The simplest instance beyond (\ref{chy1}) involves two pions and three bi-adjoint scalars,
\begin{align}
\frac{(\te{Pf} \, A_{12})^2}{z_{34} z_{45} z_{53} }&=
 \frac{ (k_{1}\cdot k_{2}) (k_{12}\cdot k_{3}) }{ (1,2,3,4,5)}
 + \frac{ (k_{1}\cdot k_{3}) (k_{13}\cdot k_{2}) }{ (1,3,2,4,5)}
 + \frac{ (k_{1}\cdot k_{3}) (k_{134}\cdot k_{2}) }{ (1,3,4,2,5) }
 \notag \\
& \! \! \! \! \! \! \! \! \! \! \! \! \! \! \! \! \! \! \! \! 
-  \frac{ (k_{1}\cdot k_{2}) (k_{12}\cdot k_{4}) }{ (1,2,4,3,5)}  
-  \frac{ (k_{1}\cdot k_{4}) (k_{14}\cdot k_{2}) }{(1,4,2,3,5)} 
-  \frac{ (k_{1}\cdot k_{4}) (k_{134}\cdot k_{2}) }{(1,4,3,2,5)} 
\ {\rm mod} \ E_i \ ,
 \label{chy9} 
\end{align}
with the underlying $W$-matrix given in (\ref{semiab68z}). While the number of terms in (\ref{chy2}) and (\ref{chy9}) 
generically grows when converting the rank-$(n{-}r)$ Pfaffians into sums over $(n{-}2)!$ permutations, our 
motivation for the rearrangement stems from the simplicity of the
Parke--Taylor form (\ref{chyPT}) for the entire $z$-dependence.


\section{Conclusions}

Here we continue the program of understanding the predictions of  color-stripped
$Z$-theory as sYM-stripped open-superstring scattering. Unlike sYM where color and kinematics, 
along with their respective Lie-algebra structures, 
can be cleanly separated, the $\ap$-dependent kinematic factors of color-stripped semi-abelian 
$Z$-theory involve functions of both CP traces and momenta.
Each of these orders in $\ap$ can be understood as part of a successive set of color-kinematic 
satisfying  effective field theories, whose culmination in $Z$-theory exhibits very soft UV behavior.  

We find compact expressions for the doubly-ordered $Z$-amplitudes whose
CP factors admit a mixture of both abelian and non-abelian generators.  At
leading order in $\ap$, these doubly-stripped amplitudes encode the predictions of a field theory 
of NLSM pions coupled to bi-adjoint scalars. Single-trace amplitudes in this theory were recently 
expressed in the CHY formalism by
Cachazo,  Cha, and Mizera~\cite{Cachazo:2016njl}.  The form of $Z$-theory's low-energy results
presented here offers an efficient complementary representation.  As color-kinematics is supported
at every order in $\ap$ (as well as the resummation), the results presented here have applicability,
through double copy, to a spectrum of theories including higher-derivative corrections to 
DBIVA$+$sYM of various supersymmetries, as well as higher-derivative corrections to scattering 
within the special-Galileon$+$NLSM$+\phi^3$ theory.  


\section*{Acknowledgements}

We are grateful to Nima Arkani-Hamed, Paolo Di Vecchia, Song He, Renata Kallosh and Radu Roiban for a combination of related collaboration and inspiring discussions, and to Freddy Cachazo, Sebastian Mizera and Guojun Zhang for encouragement and taking the time to offer helpful comments on an earlier draft.  We would like to thank IPhT at CEA-Saclay for hospitality during the initial stages of this project. OS is grateful to the University of Southampton for kind hospitality during finalization of this work. JJMC is supported by the European Research Council under ERC-STG-639729, {\it preQFT: Strategic Predictions for Quantum Field Theories}. CRM is supported by a University Research Fellowship from the Royal Society.

\appendix
\section{Expansions of semi-abelian disk integrals}
\label{app:expl}

In this appendix, we gather examples for the low-energy expansion of semi-abelian $Z$-theory amplitudes (\ref{semiab02}). The expressions can be efficiently obtained by combining their simplified $(n{-}2)!$ representation (\ref{structural}) with the Berends--Giele recursion for non-abelian $Z$-amplitudes \cite{nonabZ} (using in particular the program {\tt BGap} described in the reference). 

Not all of the examples in this appendix follow the labelling of the ${\cal W}_{\ap}$-matrices given in sections \ref{sect34} or \ref{genform} and might require some straightforward permutations of all the legs in $Z_{\Sigma}(\tau_n)$ and the momenta on the right hand side. Our subsequent choice of labels is tailored to reach all permutation-inequivalent arrangements of $\Sigma$ and $\tau_n$ from the canonical field-theory ordering $\tau_n \rightarrow \mathbb I_n \equiv 1,2,\ldots,n$.

\subsection{semi-abelian five-point integrals}

At five points, semi-abelian disk integrals $Z_\Sigma(\tau_5)$ with
three non-abelian CP factors can be found in (\ref{semiab05}). Their counterparts with four non-abelian CP factors allow
for three permutation-inequivalent arrangements of $\Sigma$ and $\tau_5$:
\begin{align}
Z_{1234}( \mathbb I_5) &=
\frac{(\pi \ap)^2}{2}  \, \Big(
 2 - \frac{ s_{45}+s_{34} }{s_{12}}- \frac{ s_{45}+s_{51} }{s_{23}} - \frac{ s_{51}+s_{12} }{s_{34}} \Big)+
           \frac{(\pi \ap)^4}{24} \,  \Big( 2s_{51}^2
          + 2s_{45}^2
\notag \\
          &\! \! \! \! \! \! \! \! \! \! \! \! \! \! \! \! \!  + 4s_{34}s_{45} + 2s_{34}^2
          + 2s_{23}s_{51}
          + 2s_{23}s_{45} + 4s_{12}s_{51}
          + 2s_{12}^2 -\frac{ s_{45}^3+2s_{34} s_{45}^2+2s_{34}^2 s_{45}+s_{34}^3  }{s_{12}} \notag \\
          &\! \! \! \! \! \! \! \! \! \! \! \! \! \! \! \! \!   - 2s_{12}s_{34}
        -\frac{ s_{45}^3+2s_{51} s_{45}^2+2s_{51}^2 s_{45}+s_{51}^3  }{s_{23}}   -\frac{ s_{12}^3+2s_{51} s_{12}^2+2s_{51}^2 s_{12}+s_{51}^3  }{s_{34}} \Big) + {\cal O}(\ap^5)
\notag
\\     
Z_{1324}( \mathbb I_5) &= 
\frac{(\pi \ap)^2}{2}  \, \frac{s_{51}+s_{45} }{s_{23}}+
     \frac{(\pi \ap)^4}{24} \, \Big(
        4s_{34}s_{45}
          + 2s_{34}^2
            - 4s_{45}s_{51}
          - 2s_{34}s_{51}
          + 2s_{23}s_{51} \notag \\
          &\! \! \! \! \! \! \! \! \! \! \! \!\! \! \! \! \!  
          + 2s_{23}s_{45}
          + 4s_{12}s_{51}
          - 2s_{12}s_{45}
          - 4s_{12}s_{34}
          + 2s_{12}^2 +\frac{   s_{51}^3
          + 2s_{45}s_{51}^2
          + 2s_{45}^2s_{51}
          + s_{45}^3}{s_{23}}
          \Big)+ {\cal O}(\ap^5) \notag
          \\
Z_{1243}( \mathbb I_5) &= \frac{(\pi\ap)^2}{2}\Big(\frac{s_{12}+s_{51}}{s_{34}}
- \frac{s_{45}+s_{34}}{s_{12}}\Big)
+ {(\pi\ap)^4\over 24}\Big( - {s_{45}^3 + 2 s_{34}s_{45}^2
+ 2 s_{34}^2 s_{45} + s_{34}^3 \over s_{12}} \notag \\
& + {s_{51}^3 + 2 s_{12}s_{51}^2 + 2 s_{12}^2 s_{51} + s_{12}^3\over
s_{34}}
	   - 2 s_{51}^2
           + 2 s_{45}^2
           +  4 s_{34} s_{51}
           + 4 s_{34} s_{45}
           + 2 s_{23} s_{51} \notag \\
          & - 2 s_{23} s_{45}
           - 4 s_{23} s_{34}
           - 4 s_{12} s_{51}
           - 4 s_{12} s_{45}
           + 4 s_{12} s_{23}\Big) + {\cal O}(\ap^5) \label{semiab06} 
\end{align}
The field-theory amplitudes along with the low-energy limit vanish
in the setup of \cite{Cachazo:2016njl} since the interaction vertices
therein do not support any couplings to an odd number of pions.


\subsection{semi-abelian six-point integrals}


Among the semi-abelian six-point integrals, any instance with $r\leq 3$ non-abelian CP factors is proportional to the NLSM amplitude by (\ref{semiab03}) and (\ref{new3g}), e.g.
\begin{align}
Z_{ijk}( \mathbb I_6) &=   \frac{1}{2} \, Z_{\times}(1,2,\ldots,6)
= \frac{ (\pi \ap)^4}{2}\,  \Big( \frac{(s_{12}+s_{23})(s_{45}+s_{56}) }{s_{123}} + 
\frac{(s_{23}+s_{34})(s_{56}+s_{61}) }{s_{234}}
 \notag \\
&\ \ \ \   + \frac{(s_{34}+s_{45})(s_{61}+s_{12}) }{s_{345}}- s_{12}- s_{23}- s_{34}- s_{45}- s_{56}- s_{61} \Big) + {\cal O}(\ap^6) \, .
\end{align}

\paragraph{Four non-abelian CP factors}

Starting with four non-abelian CP factors, one obtains inequivalent cases such as
\begin{align}
Z_{1234}( \mathbb I_6) &= (\pi \ap)^2 \Big( \frac{1}{s_{12}} \Big[ 1 - \frac{ s_{45}+s_{56} }{s_{123}}  \Big]
+ \frac{1}{s_{34}} \Big[ 1 - \frac{ s_{56}+s_{61} }{s_{234}}  \Big] \notag \\
& +\frac{1}{s_{23}} \Big[  1 -  \frac{ s_{45}+s_{56} }{s_{123}}  - \frac{ s_{56}+s_{61} }{s_{234}} \Big]  - \frac{ s_{345} + s_{56} }{s_{12}s_{34}}\Big) + {\cal O}(\ap^4)
\label{freddy}
\\
Z_{1324}( \mathbb I_6) &= \frac{   (\pi \ap)^2 }{s_{23}} \Big( \frac{ s_{45}+s_{56}}{s_{123}} + \frac{ s_{56}+s_{61} }{s_{234}} - 1 
\Big) + {\cal O}(\ap^4)
\\
Z_{1245}( \mathbb I_6) &=  (\pi \ap)^2  \Big( \frac{1}{s_{12}}+ \frac{1}{s_{45}}
- \frac{s_{123}+s_{345}}{s_{12}s_{45}}
 \Big) + {\cal O}(\ap^4) 
 \label{supp1}
\\
Z_{1254}( \mathbb I_6) &=  (\pi \ap)^2 
\Big(\frac{s_{123}+s_{345}}{s_{12}s_{45}} - \frac{1}{s_{12}}- \frac{1}{s_{45}}
 \Big)  + {\cal O}(\ap^4)
  \label{supp2}
\\
Z_{1425}( \mathbb I_6) &= 
\frac{ (\pi \ap)^4}{12}\, \Big( 3 \, \frac{ (s_{23}{+}s_{34})(s_{56}{+}s_{61}) }{s_{234}} 
+2 s_{36}
 \Big) + {\cal O}(\ap^5)
\label{except}
\\
Z_{1235}( \mathbb I_6) &=  (\pi \ap)^2  \Big( \frac{1}{s_{12} } + \frac{1}{s_{23} }  \Big) \Big( 1 - \frac{ s_{45}+s_{56}}{s_{123}}  \Big)
  + {\cal O}(\ap^4)
\\
Z_{1325}( \mathbb I_6) &= \frac{ (\pi \ap)^2 }{s_{23}} \Big( \frac{ s_{45}+s_{56}}{s_{123}} -1 \Big) + {\cal O}(\ap^4) \ .
\label{endfreddy}
\end{align}
Apart from the exception (\ref{except}) mentioned in section \ref{sect37}, the leading low-energy behaviour ties in with table (\ref{tableorder}), and a cyclic permutation of
(\ref{freddy}) matches the doubly partial amplitude
\begin{align}
A_6(1^\Sigma,2^\Sigma,3^\phi,4^\phi,5^\phi,6^\phi) &= \frac{1}{s_{56}} \Big( 1 - \frac{ s_{12}+s_{23} }{s_{123}} \Big) + \frac{1}{s_{34}} \Big( 1 - \frac{ s_{12}+s_{16} }{s_{612}} \Big) \notag \\
&-\frac{ s_{12}+s_{234} }{s_{34} s_{56}} + \frac{1}{s_{45}} \Big( 1 - \frac{ s_{12}+s_{23} }{s_{123}}  -  \frac{ s_{12}+s_{16} }{s_{612}}  \Big) 
\label{1604c}
\end{align}
involving two pions $\Sigma$ and four bi-adjoint scalars $\phi^3$ in section 3.2 of \cite{Cachazo:2016njl}. Note that $Z_{1245}(\tau_6) $ differs from $- Z_{1254}(\tau_6)$ at the orders $\alpha'^{\geq 4}$ suppressed in (\ref{supp1}) and (\ref{supp2}).


\paragraph{Five non-abelian CP factors}

With five non-abelian CP factors, we have eight inequivalent six-point cases
\begin{align}
Z_{12345}( \mathbb I_6) &= \frac{ (\pi \ap)^2}{2}  \,\Big(  \frac{1}{s_{12}} \Big[ 2 - \frac{ s_{45}+s_{56}}{s_{456} } \Big] + \frac{1}{s_{23}} \Big[ 2 - \frac{ s_{45}+s_{56}}{s_{456} }- \frac{ s_{56}+s_{61}}{s_{561} } \Big] \notag \\
&+\frac{1}{s_{34}} \Big[ 2- \frac{ s_{56}+s_{61}}{s_{561} }   - \frac{ s_{61}+s_{12}}{s_{612} }\Big] 
+\frac{1}{s_{45}} \Big[ 2  - \frac{ s_{61}+s_{12}}{s_{612} }\Big] \notag \\
&- \frac{ s_{345}+s_{56} }{s_{12}s_{34}} - \frac{ s_{61} +s_{123}}{s_{23}s_{45}} - \frac{ s_{123}+s_{345} }{s_{12}s_{45}} 
\Big) + {\cal O}(\ap^4) \label{65a} \\
Z_{12435}( \mathbb I_6) &= \frac{ (\pi \ap)^2}{2}  \, \Big(  \frac{ s_{345}+s_{56} }{s_{12}s_{34}} 
+\frac{1}{s_{34}} \Big[ \frac{ s_{56}+s_{61}}{s_{561}} + \frac{ s_{61}+s_{12}}{s_{612}}  -2\Big]\Big) + {\cal O}(\ap^4)\\
Z_{13425}( \mathbb I_6) &= \frac{ (\pi \ap)^2}{2}  \, \frac{ s_{56}+s_{61}}{s_{34} s_{561}}+ {\cal O}(\ap^4)\\
Z_{14325}( \mathbb I_6) &= - \frac{ (\pi \ap)^2}{2}  \,\Big(  \frac{1}{s_{23}}+\frac{1}{s_{34}} \Big) \, \frac{ s_{56}+s_{61}}{ s_{561}}+ {\cal O}(\ap^4) \\
Z_{13524}( \mathbb I_6) &=
\frac{ (\pi \ap)^4}{12}  \, (2 s_{45}+2s_{12} + s_{56}+s_{61}- 2 s_{123}-2 s_{345})
+ {\cal O}(\ap^5) \label{except2} \\
Z_{13254}( \mathbb I_6) &= \frac{ (\pi \ap)^2}{2}  \, \Big( \frac{ s_{45}+s_{56} }{s_{23} s_{123}} - \frac{ s_{61}+s_{123}}{s_{23} s_{45}} \Big)
+ {\cal O}(\ap^4) \\
Z_{12453}( \mathbb I_6) &= \frac{ (\pi \ap)^2}{2}  \, \Big( \frac{ s_{45}+s_{56} }{s_{12}s_{123}} + \frac{ s_{61}+s_{12}}{s_{45}s_{345}} - \frac{ s_{123} + s_{345}}{s_{12}s_{45}} \Big)
+ {\cal O}(\ap^4) \\
Z_{12354}( \mathbb I_6) &= \frac{ (\pi \ap)^2}{2}  \, \Big( \frac{1}{s_{45}} \Big[ \frac{ s_{61}+s_{12}}{s_{345}} + \frac{ s_{123}+s_{345} }{s_{12}} + \frac{ s_{61}+ s_{123} }{s_{23}} - 2 \Big]\notag \\
& -  \frac{ s_{45}+s_{56} }{ s_{123}} \Big[ \frac{1}{s_{12}}+ \frac{1}{s_{23}} \Big]  \Big)
+ {\cal O}(\ap^4) \ ,
\label{65b}
\end{align}
where also (\ref{except2}) starts at higher order $(\pi \ap)^4$ as compared to the generic expectation from table (\ref{tableorder}). Again, the low-energy limits do not have any counterparts in \cite{Cachazo:2016njl} by the odd number of NLSM pions.


\subsection{semi-abelian seven-point integrals}

The low-energy limit of the seven-point integral with $r=3$ non-abelian CP factors
\begin{align}
&\!\!\!\!\! Z_{123}(\mathbb I_7) = -( \pi \ap)^4 \Big(\frac{1}{s_{12}} \Big[ \frac{(s_{34}+s_{45})(s_{67}+s_{712})}{s_{345}} +\frac{(s_{45}+s_{56})(s_{712}+s_{123})}{s_{456}} \notag \\
&\ \ \ \ \ + \frac{(s_{56}+s_{67})(s_{123}+s_{34})}{s_{567}} 
- s_{34}-s_{45}-s_{56}-s_{67}-s_{712}-s_{123}  \Big] \notag \\
&+ \frac{1}{s_{23}} \Big[ \frac{(s_{45}+s_{56})(s_{71}+s_{123})}{s_{456}} +\frac{(s_{56}+s_{67})(s_{123}+s_{234})}{s_{567}} \label{freddy7} \\
&\ \ \ \ \ + \frac{(s_{67}+s_{71})(s_{234}+s_{45})}{s_{671}} 
- s_{45}-s_{56}-s_{67}-s_{71}-s_{123}-s_{234}  \Big] \notag \\
&+ \frac{ (s_{67}+s_{71}) (s_{34}+s_{45})}{s_{345} s_{671}} - \frac{s_{34}+s_{45}}{s_{345}}
- \frac{s_{45}+s_{56}}{s_{456}}- \frac{s_{56}+s_{67}}{s_{567}}- \frac{s_{67}+s_{71}}{s_{671}} + 2
\Big) + {\cal O}(\ap^6) \ .
\notag
\end{align}
agrees with the expression for $A_7(1^\phi,2^\phi,3^\phi,4^\Sigma,5^\Sigma,6^\Sigma,7^\Sigma)$ given in (2.25) of \cite{Cachazo:2016njl}. Moreover, there are three additional cases which cannot be obtained from relabellings of (\ref{freddy7}):
\begin{align}
&\!\!\!\!\! Z_{124}(\mathbb I_7) = ( \pi \ap)^4 \Big(\frac{1}{s_{12}} \Big[ \frac{(s_{34}+s_{45})(s_{67}+s_{712})}{s_{345}} +\frac{(s_{45}+s_{56})(s_{712}+s_{123})}{s_{456}} \notag \\
&\ \ \ \ \ + \frac{(s_{56}+s_{67})(s_{123}+s_{34})}{s_{567}} 
- s_{34}-s_{45}-s_{56}-s_{67}-s_{712}-s_{123}  \Big]  \notag \\
&+ \frac{ (s_{67}+s_{71}) (s_{34}+s_{45})}{s_{345} s_{671}} - \frac{s_{34}+s_{45}}{s_{345}}
- \frac{s_{45}+s_{56}}{s_{456}}- \frac{s_{56}+s_{67}}{s_{567}}- \frac{s_{67}+s_{71}}{s_{671}} + 2
\Big) + {\cal O}(\ap^6)  \notag
\\
&\!\!\!\!\! Z_{125}(\mathbb I_7) = ( \pi \ap)^4 \Big(\frac{1}{s_{12}} \Big[ \frac{(s_{34}+s_{45})(s_{67}+s_{712})}{s_{345}} +\frac{(s_{45}+s_{56})(s_{712}+s_{123})}{s_{456}} \notag \\
&\ \ \ \ \ + \frac{(s_{56}+s_{67})(s_{123}+s_{34})}{s_{567}} 
- s_{34}-s_{45}-s_{56}-s_{67}-s_{712}-s_{123}  \Big]  \label{freddy7b}  \\
&+ \frac{ (s_{67}+s_{71}) (s_{34}+s_{45})}{s_{345} s_{671}} 
+ \frac{ (s_{67}+s_{71}) (s_{23}+s_{34})}{s_{234} s_{671}} 
+ \frac{ (s_{23}+s_{34}) (s_{56}+s_{67})}{s_{234} s_{567}}  \notag \\
&- \frac{s_{23}+s_{34}}{s_{234}}- \frac{s_{34}+s_{45}}{s_{345}}
- \frac{s_{45}+s_{56}}{s_{456}}- \frac{s_{56}+s_{67}}{s_{567}}- \frac{s_{67}+s_{71}}{s_{671}} + 2
\Big) + {\cal O}(\ap^6)  \notag
\\
&\!\!\!\!\! Z_{135}(\mathbb I_7) = ( \pi \ap)^4 \Big(  2 - \frac{s_{23}+s_{34}}{s_{234}}
- \frac{s_{45}+s_{56}}{s_{456}}- \frac{s_{56}+s_{67}}{s_{567}}- \frac{s_{67}+s_{71}}{s_{671}}
- \frac{s_{71}+s_{12}}{s_{712}} \notag \\
&+\frac{ (s_{67}+s_{71}) (s_{23}+s_{34})}{s_{234} s_{671}} 
+ \frac{ (s_{23}+s_{34}) (s_{56}+s_{67})}{s_{234} s_{567}} + \frac{ (s_{45}+s_{56}) (s_{71}+s_{12})}{s_{456} s_{712}}  
\Big) + {\cal O}(\ap^6)  \ .
\notag
\end{align}
For selected examples with $r \geq 4$ non-abelian CP factors, the low-energy expansion starts at higher orders as compared to table (\ref{tableorder}) such as
\begin{align}
Z_{14725}(\mathbb I_7) &= - \pi^2 \ap^5 \zeta_3 s_{36} + {\cal O}(\ap^6)  \notag \\
Z_{135724}(\mathbb I_7) &=   -   \frac{ (\pi \ap)^4}{12}
\, \frac{ s_{56}+s_{67}}{s_{567}} +  \pi^2 \ap^5 \zeta_3 (s_{56}+s_{67}) \label{7shock} \\
&-\frac{  \pi^2 \ap^5 \zeta_3 }{2 s_{567}}\,  (s_{67}+s_{56})(s_{234} + s_{123} +s_{12}+s_{23}+s_{34}) + {\cal O}(\ap^6) \ . \notag     
\end{align}

\bibliographystyle{JHEP}
\bibliography{cites}{}

\providecommand{\href}[2]{#2}\begingroup\raggedright\begin{thebibliography}{10}

\bibitem{abZtheory}
J.~J.~M. Carrasco, C.~R. Mafra and O.~Schlotterer, \emph{{Abelian Z-theory:
  NLSM amplitudes and $\alpha$'-corrections from the open string}},
  \href{http://dx.doi.org/10.1007/JHEP06(2017)093}{\emph{JHEP} {\bf 06} (2017)
  093}, [\href{http://arxiv.org/abs/1608.02569}{{\tt 1608.02569}}].

\bibitem{nonabZ}
C.~R. Mafra and O.~Schlotterer, \emph{{Non-abelian $Z$-theory: Berends-Giele
  recursion for the $\alpha'$-expansion of disk integrals}},
  \href{http://dx.doi.org/10.1007/JHEP01(2017)031}{\emph{JHEP} {\bf 01} (2017)
  031}, [\href{http://arxiv.org/abs/1609.07078}{{\tt 1609.07078}}].

\bibitem{Kawai:1985xq}
H.~Kawai, D.~C. Lewellen and S.~H.~H. Tye, \emph{{A Relation Between Tree
  Amplitudes of Closed and Open Strings}},
  \href{http://dx.doi.org/10.1016/0550-3213(86)90362-7}{\emph{Nucl. Phys.} {\bf
  B269} (1986) 1--23}.

\bibitem{BCJ}
Z.~Bern, J.~J.~M. Carrasco and H.~Johansson, \emph{{New Relations for
  Gauge-Theory Amplitudes}},
  \href{http://dx.doi.org/10.1103/PhysRevD.78.085011}{\emph{Phys. Rev.} {\bf
  D78} (2008) 085011}, [\href{http://arxiv.org/abs/0805.3993}{{\tt
  0805.3993}}].

\bibitem{Cachazo:2013gna}
F.~Cachazo, S.~He and E.~Y. Yuan, \emph{{Scattering equations and
  Kawai-Lewellen-Tye orthogonality}},
  \href{http://dx.doi.org/10.1103/PhysRevD.90.065001}{\emph{Phys. Rev.} {\bf
  D90} (2014) 065001}, [\href{http://arxiv.org/abs/1306.6575}{{\tt
  1306.6575}}].

\bibitem{Brink:1976bc}
L.~Brink, J.~H. Schwarz and J.~Scherk, \emph{{Supersymmetric Yang-Mills
  Theories}}, \href{http://dx.doi.org/10.1016/0550-3213(77)90328-5}{\emph{Nucl.
  Phys.} {\bf B121} (1977) 77--92}.

\bibitem{Zfunctions}
J.~Broedel, O.~Schlotterer and S.~Stieberger, \emph{{Polylogarithms, Multiple
  Zeta Values and Superstring Amplitudes}},
  \href{http://dx.doi.org/10.1002/prop.201300019}{\emph{Fortsch. Phys.} {\bf
  61} (2013) 812--870}, [\href{http://arxiv.org/abs/1304.7267}{{\tt
  1304.7267}}].

\bibitem{Mafra:2011nv}
C.~R. Mafra, O.~Schlotterer and S.~Stieberger, \emph{{Complete N-Point
  Superstring Disk Amplitude I. Pure Spinor Computation}},
  \href{http://dx.doi.org/10.1016/j.nuclphysb.2013.04.023}{\emph{Nucl. Phys.}
  {\bf B873} (2013) 419--460}, [\href{http://arxiv.org/abs/1106.2645}{{\tt
  1106.2645}}].

\bibitem{Mafra:2011nw}
C.~R. Mafra, O.~Schlotterer and S.~Stieberger, \emph{{Complete N-Point
  Superstring Disk Amplitude II. Amplitude and Hypergeometric Function
  Structure}},
  \href{http://dx.doi.org/10.1016/j.nuclphysb.2013.04.022}{\emph{Nucl. Phys.}
  {\bf B873} (2013) 461--513}, [\href{http://arxiv.org/abs/1106.2646}{{\tt
  1106.2646}}].

\bibitem{Bern:1999bx}
Z.~Bern, A.~De~Freitas and H.~L. Wong, \emph{{On the coupling of gravitons to
  matter}}, \href{http://dx.doi.org/10.1103/PhysRevLett.84.3531}{\emph{Phys.
  Rev. Lett.} {\bf 84} (2000) 3531},
  [\href{http://arxiv.org/abs/hep-th/9912033}{{\tt hep-th/9912033}}].

\bibitem{Cachazo:2013iea}
F.~Cachazo, S.~He and E.~Y. Yuan, \emph{{Scattering of Massless Particles:
  Scalars, Gluons and Gravitons}},
  \href{http://dx.doi.org/10.1007/JHEP07(2014)033}{\emph{JHEP} {\bf 07} (2014)
  033}, [\href{http://arxiv.org/abs/1309.0885}{{\tt 1309.0885}}].

\bibitem{Cronin:1967jq}
J.~A. Cronin, \emph{{Phenomenological model of strong and weak interactions in
  chiral U(3) x U(3)}},
  \href{http://dx.doi.org/10.1103/PhysRev.161.1483}{\emph{Phys. Rev.} {\bf 161}
  (1967) 1483--1494}.

\bibitem{Weinberg:1966fm}
S.~Weinberg, \emph{{Dynamical approach to current algebra}},
  \href{http://dx.doi.org/10.1103/PhysRevLett.18.188}{\emph{Phys. Rev. Lett.}
  {\bf 18} (1967) 188--191}.

\bibitem{Weinberg:1968de}
S.~Weinberg, \emph{{Nonlinear realizations of chiral symmetry}},
  \href{http://dx.doi.org/10.1103/PhysRev.166.1568}{\emph{Phys. Rev.} {\bf 166}
  (1968) 1568--1577}.

\bibitem{Brown:1967qh}
L.~S. Brown, \emph{{Field theory of chiral symmetry}},
  \href{http://dx.doi.org/10.1103/PhysRev.163.1802}{\emph{Phys. Rev.} {\bf 163}
  (1967) 1802--1807}.

\bibitem{Chang:1967zza}
P.~Chang and F.~Gursey, \emph{{Unified Formulation of Effective Nonlinear
  Pion-Nucleon Lagrangians}},
  \href{http://dx.doi.org/10.1103/PhysRev.164.1752}{\emph{Phys. Rev.} {\bf 164}
  (1967) 1752--1761}.

\bibitem{Susskind:1970gf}
L.~Susskind and G.~Frye, \emph{{Algebraic aspects of pionic duality diagrams}},
  \href{http://dx.doi.org/10.1103/PhysRevD.1.1682}{\emph{Phys. Rev.} {\bf D1}
  (1970) 1682--1686}.

\bibitem{Osborn:1969ku}
H.~Osborn, \emph{{Implications of adler zeros for multipion processes}},
  \href{http://dx.doi.org/10.1007/BF02755724}{\emph{Lett. Nuovo Cim.} {\bf 2S1}
  (1969) 717--723}.

\bibitem{Ellis:1970nt}
J.~R. Ellis and B.~Renner, \emph{{On the relationship between chiral and dual
  models}}, \href{http://dx.doi.org/10.1016/0550-3213(70)90515-8}{\emph{Nucl.
  Phys.} {\bf B21} (1970) 205--216}.

\bibitem{Kampf:2013vha}
K.~Kampf, J.~Novotny and J.~Trnka, \emph{{Tree-level Amplitudes in the
  Nonlinear Sigma Model}},
  \href{http://dx.doi.org/10.1007/JHEP05(2013)032}{\emph{JHEP} {\bf 05} (2013)
  032}, [\href{http://arxiv.org/abs/1304.3048}{{\tt 1304.3048}}].

\bibitem{Gross:1987kza}
D.~J. Gross and P.~F. Mende, \emph{{The High-Energy Behavior of String
  Scattering Amplitudes}},
  \href{http://dx.doi.org/10.1016/0370-2693(87)90355-8}{\emph{Phys. Lett.} {\bf
  B197} (1987) 129--134}.

\bibitem{Caron-Huot:2016icg}
S.~Caron-Huot, Z.~Komargodski, A.~Sever and A.~Zhiboedov, \emph{{Strings from
  Massive Higher Spins: The Asymptotic Uniqueness of the Veneziano Amplitude}},
  \href{http://dx.doi.org/10.1007/JHEP10(2017)026}{\emph{JHEP} {\bf 10} (2017)
  026}, [\href{http://arxiv.org/abs/1607.04253}{{\tt 1607.04253}}].

\bibitem{Arkani-HamedStrings2016}
N.~Arkani-Hamed and Y.-T. Huang, ``{Towards deriving String Theory as the
  weakly coupled UV-completion of Gravity}.''
  \url{http://ymsc.tsinghua.edu.cn:8090/strings/slides/8.3/Nima\%208.3.pdf}.

\bibitem{Cachazo:2016njl}
F.~Cachazo, P.~Cha and S.~Mizera, \emph{{Extensions of Theories from Soft
  Limits}}, \href{http://dx.doi.org/10.1007/JHEP06(2016)170}{\emph{JHEP} {\bf
  06} (2016) 170}, [\href{http://arxiv.org/abs/1604.03893}{{\tt 1604.03893}}].

\bibitem{Cachazo:2013hca}
F.~Cachazo, S.~He and E.~Y. Yuan, \emph{{Scattering of Massless Particles in
  Arbitrary Dimensions}},
  \href{http://dx.doi.org/10.1103/PhysRevLett.113.171601}{\emph{Phys. Rev.
  Lett.} {\bf 113} (2014) 171601}, [\href{http://arxiv.org/abs/1307.2199}{{\tt
  1307.2199}}].

\bibitem{Cachazo:2014xea}
F.~Cachazo, S.~He and E.~Y. Yuan, \emph{{Scattering Equations and Matrices:
  From Einstein To Yang-Mills, DBI and NLSM}},
  \href{http://dx.doi.org/10.1007/JHEP07(2015)149}{\emph{JHEP} {\bf 07} (2015)
  149}, [\href{http://arxiv.org/abs/1412.3479}{{\tt 1412.3479}}].

\bibitem{Bergshoeff:2013pia}
E.~Bergshoeff, F.~Coomans, R.~Kallosh, C.~S. Shahbazi and A.~Van~Proeyen,
  \emph{{Dirac-Born-Infeld-Volkov-Akulov and Deformation of Supersymmetry}},
  \href{http://dx.doi.org/10.1007/JHEP08(2013)100}{\emph{JHEP} {\bf 08} (2013)
  100}, [\href{http://arxiv.org/abs/1303.5662}{{\tt 1303.5662}}].

\bibitem{momentumKernel}
N.~E.~J. Bjerrum-Bohr, P.~H. Damgaard, T.~Sondergaard and P.~Vanhove,
  \emph{{The Momentum Kernel of Gauge and Gravity Theories}},
  \href{http://dx.doi.org/10.1007/JHEP01(2011)001}{\emph{JHEP} {\bf 01} (2011)
  001}, [\href{http://arxiv.org/abs/1010.3933}{{\tt 1010.3933}}].

\bibitem{loopBCJ}
Z.~Bern, J.~J.~M. Carrasco and H.~Johansson, \emph{{Perturbative Quantum
  Gravity as a Double Copy of Gauge Theory}},
  \href{http://dx.doi.org/10.1103/PhysRevLett.105.061602}{\emph{Phys. Rev.
  Lett.} {\bf 105} (2010) 061602}, [\href{http://arxiv.org/abs/1004.0476}{{\tt
  1004.0476}}].

\bibitem{Elvang:2013cua}
H.~Elvang and Y.-t. Huang, \emph{{Scattering Amplitudes}},
  \href{http://arxiv.org/abs/1308.1697}{{\tt 1308.1697}}.

\bibitem{Ferrara:2014kva}
S.~Ferrara, R.~Kallosh and A.~Linde, \emph{{Cosmology with Nilpotent
  Superfields}}, \href{http://dx.doi.org/10.1007/JHEP10(2014)143}{\emph{JHEP}
  {\bf 10} (2014) 143}, [\href{http://arxiv.org/abs/1408.4096}{{\tt
  1408.4096}}].

\bibitem{Ferrara:2015tyn}
S.~Ferrara, R.~Kallosh and J.~Thaler, \emph{{Cosmology with orthogonal
  nilpotent superfields}},
  \href{http://dx.doi.org/10.1103/PhysRevD.93.043516}{\emph{Phys. Rev.} {\bf
  D93} (2016) 043516}, [\href{http://arxiv.org/abs/1512.00545}{{\tt
  1512.00545}}].

\bibitem{Carrasco:2015pla}
J.~J.~M. Carrasco, R.~Kallosh and A.~Linde, \emph{$\alpha $-attractors: Planck,
  lhc and dark energy},
  \href{http://dx.doi.org/10.1007/JHEP10(2015)147}{\emph{JHEP} {\bf 10} (2015)
  147}, [\href{http://arxiv.org/abs/1506.01708}{{\tt 1506.01708}}].

\bibitem{Carrasco:2015iij}
J.~J.~M. Carrasco, R.~Kallosh and A.~Linde, \emph{{Minimal supergravity
  inflation}}, \href{http://dx.doi.org/10.1103/PhysRevD.93.061301}{\emph{Phys.
  Rev.} {\bf D93} (2016) 061301}, [\href{http://arxiv.org/abs/1512.00546}{{\tt
  1512.00546}}].

\bibitem{Bergshoeff:2015tra}
E.~A. Bergshoeff, D.~Z. Freedman, R.~Kallosh and A.~Van~Proeyen, \emph{{Pure de
  Sitter Supergravity}}, \href{http://dx.doi.org/10.1103/PhysRevD.93.069901,
  10.1103/PhysRevD.92.085040}{\emph{Phys. Rev.} {\bf D92} (2015) 085040},
  [\href{http://arxiv.org/abs/1507.08264}{{\tt 1507.08264}}].

\bibitem{Hasegawa:2015bza}
F.~Hasegawa and Y.~Yamada, \emph{{Component action of nilpotent multiplet
  coupled to matter in 4 dimensional $ \mathcal{N}=1 $ supergravity}},
  \href{http://dx.doi.org/10.1007/JHEP10(2015)106}{\emph{JHEP} {\bf 10} (2015)
  106}, [\href{http://arxiv.org/abs/1507.08619}{{\tt 1507.08619}}].

\bibitem{Kuzenko:2015yxa}
S.~M. Kuzenko, \emph{{Complex linear Goldstino superfield and supergravity}},
  \href{http://dx.doi.org/10.1007/JHEP10(2015)006}{\emph{JHEP} {\bf 10} (2015)
  006}, [\href{http://arxiv.org/abs/1508.03190}{{\tt 1508.03190}}].

\bibitem{Bandos:2015xnf}
I.~Bandos, L.~Martucci, D.~Sorokin and M.~Tonin, \emph{{Brane induced
  supersymmetry breaking and de Sitter supergravity}},
  \href{http://dx.doi.org/10.1007/JHEP02(2016)080}{\emph{JHEP} {\bf 02} (2016)
  080}, [\href{http://arxiv.org/abs/1511.03024}{{\tt 1511.03024}}].

\bibitem{Adler:1964um}
S.~L. Adler, \emph{{Consistency conditions on the strong interactions implied
  by a partially conserved axial vector current}},
  \href{http://dx.doi.org/10.1103/PhysRev.137.B1022}{\emph{Phys. Rev.} {\bf
  137} (1965) B1022--B1033}.

\bibitem{ArkaniHamed:2008gz}
N.~Arkani-Hamed, F.~Cachazo and J.~Kaplan, \emph{{What is the Simplest Quantum
  Field Theory?}}, \href{http://dx.doi.org/10.1007/JHEP09(2010)016}{\emph{JHEP}
  {\bf 09} (2010) 016}, [\href{http://arxiv.org/abs/0808.1446}{{\tt
  0808.1446}}].

\bibitem{Kallosh:2016qvo}
R.~Kallosh, \emph{{Nonlinear (Super)Symmetries and Amplitudes}},
  \href{http://dx.doi.org/10.1007/JHEP03(2017)038}{\emph{JHEP} {\bf 03} (2017)
  038}, [\href{http://arxiv.org/abs/1609.09123}{{\tt 1609.09123}}].

\bibitem{Kallosh:2016lwj}
R.~Kallosh, A.~Karlsson and D.~Murli, \emph{{Origin of Soft Limits from
  Nonlinear Supersymmetry in Volkov-Akulov Theory}},
  \href{http://dx.doi.org/10.1007/JHEP03(2017)081}{\emph{JHEP} {\bf 03} (2017)
  081}, [\href{http://arxiv.org/abs/1609.09127}{{\tt 1609.09127}}].

\bibitem{Cheung:2016drk}
C.~Cheung, K.~Kampf, J.~Novotny, C.-H. Shen and J.~Trnka, \emph{{A Periodic
  Table of Effective Field Theories}},
  \href{http://dx.doi.org/10.1007/JHEP02(2017)020}{\emph{JHEP} {\bf 02} (2017)
  020}, [\href{http://arxiv.org/abs/1611.03137}{{\tt 1611.03137}}].

\bibitem{Du:2016njc}
Y.-J. Du and H.~Luo, \emph{{Leading order multi-soft behaviors of tree
  amplitudes in NLSM}},
  \href{http://dx.doi.org/10.1007/JHEP03(2017)062}{\emph{JHEP} {\bf 03} (2017)
  062}, [\href{http://arxiv.org/abs/1611.07479}{{\tt 1611.07479}}].

\bibitem{BjerrumBohr:2009rd}
N.~E.~J. Bjerrum-Bohr, P.~H. Damgaard and P.~Vanhove, \emph{{Minimal Basis for
  Gauge Theory Amplitudes}},
  \href{http://dx.doi.org/10.1103/PhysRevLett.103.161602}{\emph{Phys. Rev.
  Lett.} {\bf 103} (2009) 161602}, [\href{http://arxiv.org/abs/0907.1425}{{\tt
  0907.1425}}].

\bibitem{Stieberger:2009hq}
S.~Stieberger, \emph{{Open \& Closed vs. Pure Open String Disk Amplitudes}},
  \href{http://arxiv.org/abs/0907.2211}{{\tt 0907.2211}}.

\bibitem{goodCDOpenString}
Y.-X. Chen, Y.-J. Du and Q.~Ma, \emph{{On Primary Relations at Tree-level in
  String Theory and Field Theory}},
  \href{http://dx.doi.org/10.1007/JHEP02(2012)061}{\emph{JHEP} {\bf 02} (2012)
  061}, [\href{http://arxiv.org/abs/1109.0685}{{\tt 1109.0685}}].

\bibitem{Mafra:2016ltu}
C.~R. Mafra, \emph{{Berends-Giele recursion for double-color-ordered
  amplitudes}}, \href{http://dx.doi.org/10.1007/JHEP07(2016)080}{\emph{JHEP}
  {\bf 07} (2016) 080}, [\href{http://arxiv.org/abs/1603.09731}{{\tt
  1603.09731}}].

\bibitem{Bern:1998sv}
Z.~Bern, L.~J. Dixon, M.~Perelstein and J.~S. Rozowsky, \emph{{Multileg one
  loop gravity amplitudes from gauge theory}},
  \href{http://dx.doi.org/10.1016/S0550-3213(99)00029-2}{\emph{Nucl. Phys.}
  {\bf B546} (1999) 423--479}, [\href{http://arxiv.org/abs/hep-th/9811140}{{\tt
  hep-th/9811140}}].

\bibitem{Metsaev:1987qp}
R.~R. Metsaev, M.~Rakhmanov and A.~A. Tseytlin, \emph{{The {Born-Infeld} Action
  as the Effective Action in the Open Superstring Theory}},
  \href{http://dx.doi.org/10.1016/0370-2693(87)91223-8}{\emph{Phys. Lett.} {\bf
  B193} (1987) 207--212}.

\bibitem{Schlotterer:2012ny}
O.~Schlotterer and S.~Stieberger, \emph{{Motivic Multiple Zeta Values and
  Superstring Amplitudes}},
  \href{http://dx.doi.org/10.1088/1751-8113/46/47/475401}{\emph{J. Phys.} {\bf
  A46} (2013) 475401}, [\href{http://arxiv.org/abs/1205.1516}{{\tt
  1205.1516}}].

\bibitem{Broedel:2013aza}
J.~Broedel, O.~Schlotterer, S.~Stieberger and T.~Terasoma, \emph{{All order
  $\alpha^{\prime}$-expansion of superstring trees from the Drinfeld
  associator}}, \href{http://dx.doi.org/10.1103/PhysRevD.89.066014}{\emph{Phys.
  Rev.} {\bf D89} (2014) 066014}, [\href{http://arxiv.org/abs/1304.7304}{{\tt
  1304.7304}}].

\bibitem{DelDuca:1999rs}
V.~Del~Duca, L.~J. Dixon and F.~Maltoni, \emph{{New color decompositions for
  gauge amplitudes at tree and loop level}},
  \href{http://dx.doi.org/10.1016/S0550-3213(99)00809-3}{\emph{Nucl. Phys.}
  {\bf B571} (2000) 51--70}, [\href{http://arxiv.org/abs/hep-ph/9910563}{{\tt
  hep-ph/9910563}}].

\bibitem{ckNLSM}
Y.-J. Du and C.-H. Fu, \emph{{Explicit BCJ numerators of nonlinear simga
  model}}, \href{http://dx.doi.org/10.1007/JHEP09(2016)174}{\emph{JHEP} {\bf
  09} (2016) 174}, [\href{http://arxiv.org/abs/1606.05846}{{\tt 1606.05846}}].

\bibitem{Mafra:2011kj}
C.~R. Mafra, O.~Schlotterer and S.~Stieberger, \emph{{Explicit BCJ Numerators
  from Pure Spinors}},
  \href{http://dx.doi.org/10.1007/JHEP07(2011)092}{\emph{JHEP} {\bf 07} (2011)
  092}, [\href{http://arxiv.org/abs/1104.5224}{{\tt 1104.5224}}].

\bibitem{Cheung:2016prv}
C.~Cheung and C.-H. Shen, \emph{{Symmetry for Flavor-Kinematics Duality from an
  Action}}, \href{http://dx.doi.org/10.1103/PhysRevLett.118.121601}{\emph{Phys.
  Rev. Lett.} {\bf 118} (2017) 121601},
  [\href{http://arxiv.org/abs/1612.00868}{{\tt 1612.00868}}].

\bibitem{Medina:2002nk}
R.~Medina, F.~T. Brandt and F.~R. Machado, \emph{{The Open superstring five
  point amplitude revisited}},
  \href{http://dx.doi.org/10.1088/1126-6708/2002/07/071}{\emph{JHEP} {\bf 07}
  (2002) 071}, [\href{http://arxiv.org/abs/hep-th/0208121}{{\tt
  hep-th/0208121}}].

\bibitem{Barreiro:2005hv}
L.~A. Barreiro and R.~Medina, \emph{{5-field terms in the open superstring
  effective action}},
  \href{http://dx.doi.org/10.1088/1126-6708/2005/03/055}{\emph{JHEP} {\bf 03}
  (2005) 055}, [\href{http://arxiv.org/abs/hep-th/0503182}{{\tt
  hep-th/0503182}}].

\bibitem{Oprisa:2005wu}
D.~Oprisa and S.~Stieberger, \emph{{Six gluon open superstring disk amplitude,
  multiple hypergeometric series and Euler-Zagier sums}},
  \href{http://arxiv.org/abs/hep-th/0509042}{{\tt hep-th/0509042}}.

\bibitem{Stieberger:2006te}
S.~Stieberger and T.~R. Taylor, \emph{{Multi-Gluon Scattering in Open
  Superstring Theory}},
  \href{http://dx.doi.org/10.1103/PhysRevD.74.126007}{\emph{Phys. Rev.} {\bf
  D74} (2006) 126007}, [\href{http://arxiv.org/abs/hep-th/0609175}{{\tt
  hep-th/0609175}}].

\bibitem{Boels:2013jua}
R.~H. Boels, \emph{{On the field theory expansion of superstring five point
  amplitudes}},
  \href{http://dx.doi.org/10.1016/j.nuclphysb.2013.08.009}{\emph{Nucl. Phys.}
  {\bf B876} (2013) 215--233}, [\href{http://arxiv.org/abs/1304.7918}{{\tt
  1304.7918}}].

\bibitem{Puhlfuerst:2015gta}
G.~Puhlfuerst and S.~Stieberger, \emph{{Differential Equations, Associators,
  and Recurrences for Amplitudes}},
  \href{http://dx.doi.org/10.1016/j.nuclphysb.2015.11.005}{\emph{Nucl. Phys.}
  {\bf B902} (2016) 186--245}, [\href{http://arxiv.org/abs/1507.01582}{{\tt
  1507.01582}}].

\bibitem{Aomoto}
K.~Aomoto, \emph{{Special values of hyperlogarithms and linear difference
  schemes}}, {\emph{Illinois J. Math.} {\bf 34} (1990) 191}.

\bibitem{Terasoma}
T.~Terasoma, \emph{{Selberg integrals and multiple zeta values}},
  {\emph{Compositio Mathematica} {\bf 133} (2002) 1}.

\bibitem{Brown:2009qja}
F.~C.~S. Brown, \emph{{Multiple zeta values and periods of moduli spaces ${\cal
  M}_{0 ,n}( \mathbb R )$}}, {\emph{Annales Sci. Ecole Norm. Sup.} {\bf 42}
  (2009) 371}, [\href{http://arxiv.org/abs/math/0606419}{{\tt math/0606419}}].

\bibitem{Stieberger:2009rr}
S.~Stieberger, \emph{{Constraints on Tree-Level Higher Order Gravitational
  Couplings in Superstring Theory}},
  \href{http://dx.doi.org/10.1103/PhysRevLett.106.111601}{\emph{Phys. Rev.
  Lett.} {\bf 106} (2011) 111601}, [\href{http://arxiv.org/abs/0910.0180}{{\tt
  0910.0180}}].

\bibitem{Drummond:2013vz}
J.~M. Drummond and E.~Ragoucy, \emph{{Superstring amplitudes and the
  associator}}, \href{http://dx.doi.org/10.1007/JHEP08(2013)135}{\emph{JHEP}
  {\bf 08} (2013) 135}, [\href{http://arxiv.org/abs/1301.0794}{{\tt
  1301.0794}}].

\bibitem{wwwMZV}
J.~Broedel, O.~Schlotterer and S.~Stieberger. \url{http://mzv.mpp.mpg.de}.

\bibitem{Green:1995ga}
M.~B. Green and M.~Gutperle, \emph{{Symmetry breaking at enhanced symmetry
  points}}, \href{http://dx.doi.org/10.1016/0550-3213(95)00608-7}{\emph{Nucl.
  Phys.} {\bf B460} (1996) 77--108},
  [\href{http://arxiv.org/abs/hep-th/9509171}{{\tt hep-th/9509171}}].

\bibitem{Stieberger:2016lng}
S.~Stieberger and T.~R. Taylor, \emph{{New relations for Einstein-Yang-Mills
  amplitudes}},
  \href{http://dx.doi.org/10.1016/j.nuclphysb.2016.09.014}{\emph{Nucl. Phys.}
  {\bf B913} (2016) 151--162}, [\href{http://arxiv.org/abs/1606.09616}{{\tt
  1606.09616}}].

\bibitem{Nandan:2016pya}
D.~Nandan, J.~Plefka, O.~Schlotterer and C.~Wen, \emph{{Einstein-Yang-Mills
  from pure Yang-Mills amplitudes}},
  \href{http://dx.doi.org/10.1007/JHEP10(2016)070}{\emph{JHEP} {\bf 10} (2016)
  070}, [\href{http://arxiv.org/abs/1607.05701}{{\tt 1607.05701}}].

\bibitem{delaCruz:2016gnm}
L.~de~la Cruz, A.~Kniss and S.~Weinzierl, \emph{{Relations for
  Einstein-Yang-Mills amplitudes from the CHY representation}},
  \href{http://dx.doi.org/10.1016/j.physletb.2017.01.036}{\emph{Phys. Lett.}
  {\bf B767} (2017) 86--90}, [\href{http://arxiv.org/abs/1607.06036}{{\tt
  1607.06036}}].

\bibitem{Schlotterer:2016cxa}
O.~Schlotterer, \emph{{Amplitude relations in heterotic string theory and
  Einstein-Yang-Mills}},
  \href{http://dx.doi.org/10.1007/JHEP11(2016)074}{\emph{JHEP} {\bf 11} (2016)
  074}, [\href{http://arxiv.org/abs/1608.00130}{{\tt 1608.00130}}].

\bibitem{Du:2016wkt}
Y.-J. Du, F.~Teng and Y.-S. Wu, \emph{{Direct Evaluation of $n$-point
  single-trace MHV amplitudes in 4d Einstein-Yang-Mills theory using the CHY
  Formalism}}, \href{http://dx.doi.org/10.1007/JHEP09(2016)171}{\emph{JHEP}
  {\bf 09} (2016) 171}, [\href{http://arxiv.org/abs/1608.00883}{{\tt
  1608.00883}}].

\bibitem{Cachazo:2015nwa}
F.~Cachazo and H.~Gomez, \emph{{Computation of Contour Integrals on ${\cal
  M}_{0,n}$}}, \href{http://dx.doi.org/10.1007/JHEP04(2016)108}{\emph{JHEP}
  {\bf 04} (2016) 108}, [\href{http://arxiv.org/abs/1505.03571}{{\tt
  1505.03571}}].

\bibitem{Cardona:2016gon}
C.~Cardona, B.~Feng, H.~Gomez and R.~Huang, \emph{{Cross-ratio Identities and
  Higher-order Poles of CHY-integrand}},
  \href{http://dx.doi.org/10.1007/JHEP09(2016)133}{\emph{JHEP} {\bf 09} (2016)
  133}, [\href{http://arxiv.org/abs/1606.00670}{{\tt 1606.00670}}].

\end{thebibliography}\endgroup

\end{document}